\documentclass[sigplan,screen,nonacm]{acmart}

\renewcommand\footnotetextcopyrightpermission[1]{}
\makeatletter
\newcommand{\paper@inlineauthor}[1]{%
  \ifx\@currentauthors\@empty
    \gdef\@currentauthors{\mbox{#1}}%
  \else
    \g@addto@macro\@currentauthors{\quad\mbox{#1}}%
  \fi
  \gdef\and{}%
}
\patchcmd{\@mkauthors@iii}
  {\hsize=\textwidth}
  {\let\@author\paper@inlineauthor\hsize=\textwidth}
  {}{\PackageError{paper}{Could not apply horizontal author layout}{Check the acmart author formatting.}}
\newcommand{\authoraffiliationmarks}[1]{%
  \expandafter\gdef\csname typeset@author\the\num@authors\endcsname##1{%
    ##1\textsuperscript{#1}}%
}
\makeatother

\AtBeginDocument{%
  }

\usepackage{tikz}
\definecolor{figred}{HTML}{C00000}
\definecolor{figgreen}{HTML}{789440}
\newcommand{\circnum}[2]{\tikz[baseline=(c.base)]\node[circle,fill=#1,text=white,inner sep=0pt,minimum size=1.10em,font=\sffamily\large](c){#2};}
\newcommand{\rednum}[1]{\circnum{figred}{#1}}

\setcopyright{none}
\acmYear{}
\copyrightyear{}
\acmDOI{}
\acmISBN{}

\begin{document}

\title[StageBreaker]{StageBreaker: Shifting Video RL Post-Training Scheduling from Stages down to Data Dependencies}

\author{Li Peng}
\authoraffiliationmarks{1,2}
\author{Chenghuan Huang}
\authoraffiliationmarks{2}
\author{Ziyang Ma}
\authoraffiliationmarks{2}
\author{Jing Lyu}
\authoraffiliationmarks{3}
\author{Xiang Chen}
\authoraffiliationmarks{1}
\author{Jie Zhang}
\authoraffiliationmarks{1}

\affiliation[obeypunctuation=true]{%
  \institution{\textsuperscript{1}Peking University}\quad
  \institution{\textsuperscript{2}WeChat HPC, Tencent Inc.}\quad
  \institution{\textsuperscript{3}WeChat Vision, Tencent Inc.}
  \country{}
}

\renewcommand{\shortauthors}{Li Peng et al.}

\begin{abstract}
Reinforcement learning (RL) post-training has emerged as the key paradigm for improving autoregressive video models.
Emerging vision-language reward models further strengthen post-training with accurate rewards, but their long chain-of-thought inference substantially raises the associated costs, while the pursuit of long videos prolongs policy training.
Despite these rising costs, existing video RL systems still adopt coarse-grained stage-centric execution, which stalls policy training until all reward evaluation finishes, replays every trajectory twice due to a further barrier within policy training, and induces load imbalance across GPUs.
To address the challenges, we propose StageBreaker, a fine-grained data-dependency-driven scheduling framework, which schedules the true dataflow of an RL step rather than its stages.
Specifically, StageBreaker overlaps policy training with reward evaluation by exploiting reward-separable policy training, and merges the two trajectory replays into one by leveraging loss-independent back-propagation.
It further balances GPU load by scheduling computation at the finer granularity.
Our evaluation results show that StageBreaker achieves a speedup of up to 1.41$\times$ over stage-centric execution.
\end{abstract}

\maketitle

\section{Introduction}

Autoregressive video models (AR models), which generate long videos chunk by chunk (i.e., multiple video frames), now power long-form and streaming video applications \cite{wu2025hunyuanvideo,wan2025wan,teng2025magi,chen2025skyreels,yin2025slow,huang2025selfforcing,yang2026longlive,liu2026rolling,cui2026self,chen2026longlive}.
As these models enter real-world deployment, generation quality becomes the decisive requirement, in particular alignment with human preference, aesthetic quality, and temporal coherence \cite{gao2025seedance,seedream2025seedream,lu2026raven,zhang2026astrolabe,he2026ar,wang2026tempact}. %
By delivering substantial gains across all these dimensions, reinforcement learning (RL) post-training has emerged as the key paradigm to meet this requirement~\cite{xue2025dancegrpo,liu2025flow,zhang2026kvpo}.

\begin{figure}[t]
  \centering
  \includegraphics[width=0.99\linewidth]{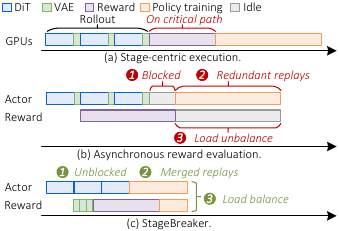}
  \vspace{-10pt}
  \caption{Comparison of RL step execution.}
  \vspace{-13pt}
  \label{fig:intro}
\end{figure}

RL realizes these gains through a training loop of repeated steps, each of which comprises rollout, reward evaluation, and policy training (cf. Figure~\ref{fig:intro}a)~\cite{sheng2025hybridflow,zhu2026distrs,mazzaglia2024genrl}.
In the rollout stage, the video model being trained (\emph{the policy}) generates multiple candidate videos for each prompt in a batch~\cite{sheng2025hybridflow,hu2025openrlhf}.
A reward model then scores each candidate on desired dimensions ~\cite{liu2025improving,wang2026unified,wang2025vr,he2025videoscore2}, and policy training replays each candidate's trajectory (i.e., its sequence of generation steps) and combines the result with reward scores to compute the loss~\cite{wang2026leangrpo,xue2025dancegrpo}.
Policy training then back-propagates this loss to update the policy, commonly through low-rank adaptation (LoRA)~\cite{zhang2026kvpo,dancegrpo_github},  raising the probability of generating high-scoring trajectories.

The continued pursuit of high-quality and long videos \cite{liu2025flow,xue2025dancegrpo,lu2026raven,zhang2026astrolabe,he2026ar} raises the cost of each RL step.
Unlike earlier lightweight reward metrics~\cite{wu2023human,wang2026unified,wang2026unifiedThink}, which cannot comprehensively and reliably assess video qualities, recent reward models employ large vision-language models (VLMs) to examine multiple video frames and perform long chain-of-thought (CoT) reasoning to produce holistic judgments for video quality~\cite{wang2025vr,liu2025improving,he2025videoscore2,wang2026cac,wang2024lift}.
Longer videos, in turn, prolong policy training, whose trajectory replay re-runs the video model for each video generation step~\cite{xue2025dancegrpo,liu2025flow}.
Together, reward evaluation and policy training now account for 63$\sim$75\% of an RL step.

These rising costs unfortunately expose a mismatch with current video RL systems.
As illustrated in Figure \ref{fig:intro}a, these systems \cite{sheng2025hybridflow,mazzaglia2024genrl,hu2025openrlhf,verl,verl_omni} organize each RL step into successive rollout, reward, and policy training stages (i.e., \emph{stage-centric execution}).
Each stage serves as the unit of computation scheduling and GPU assignment, processing the entire batch on a fixed set of GPUs before the next stage starts~\cite{sheng2025hybridflow,hu2025openrlhf,slime_github}.
However, these stage boundaries impose coarser constraints than the true data dependencies within an RL step, prolonging each step and underutilizing GPUs.
For example, while the reward model can score a candidate upon its generation, stage-centric execution starts reward evaluation only after the entire rollout stage finishes, exposing its cost entirely in the latency of each RL step.
To take this cost off the critical path, as shown in Figure \ref{fig:intro}b, recent works~\cite{verlomni_async_reward,gao2026rollpacker,wang2025reinforcement} deploy reward models on dedicated GPUs (reward GPUs) and score each candidate upon generation, leaving the remaining GPUs (actor GPUs) to run rollout and policy training.
In practice, this mechanism is only a preliminary exploration that falls short of the expected gains and even slows RL post-training by 20$\sim$29\%, as three key challenges remain unaddressed.

\emph{(1) Reward-blocked policy training.}
Existing RL systems stall policy training until the reward stage scores the entire batch, as policy training needs the complete set of scores to derive the \emph{advantages} (the relative quality of each candidate) for the subsequent loss computation~\cite{sheng2025hybridflow,liu2025flow,kvpo_github,zhu2026distrs,zhong2025optimizing}.
Even under asynchronous reward evaluation, the actor GPUs stay idle for 13\% of each RL step before starting policy training, waiting for the last reward score of the batch (\rednum{1} in Figure \ref{fig:intro}b).
\emph{(2) Redundant trajectory replay.}
As in conventional model training, existing RL systems first compute the loss over the entire batch, then back-propagate it to derive the \emph{policy gradient}, and finally update the policy.
Policy training therefore runs a forward pass for the loss, in which it replays all trajectories to derive their probabilities~\cite{zhang2026kvpo,liu2025flow}.
The backward pass, which back-propagates this loss, starts only after the forward pass finishes, imposing a further batch-level barrier within policy training.
Although the backward pass requires these probabilities as well, GPU memory cannot retain the activations of all trajectories until then.
The backward pass thus repeats the costly replay (\rednum{2}), resulting in redundant replay for 19$\sim$26\% of each RL step.
\emph{(3) Stage-induced load imbalance.}
Existing RL systems bind computation to GPUs at the coarse granularity of stages, making it difficult to balance the load evenly across GPUs (\rednum{3}).
For example, asynchronous reward evaluation, following this binding, dedicates GPUs to the reward stage but overlooks their potential to execute other computation \cite{sheng2025hybridflow,verl_omni,verlomni_async_reward}. The reward GPUs thus sit idle for 64\% of each RL step once scoring finishes, while the actor GPUs carry all rollout and policy training.

As these limitations all stem from the coarse granularity of stages, our key insight is to schedule the true dataflow, which uncovers potential to shorten each RL step and fill idle GPUs.
\emph{(1) Reward-separable policy training.}
Policy training can in fact proceed asynchronously with reward evaluation, as the loss it computes is simply the product of the advantage and a reward-independent part.
Policy training can thus compute and back-propagate the reward-independent part while reward evaluation is still in progress, and combine it with the advantages after both are complete.
\emph{(2) Loss-independent back-propagation.}
Within policy training, back-propagation can likewise proceed without waiting for the loss, as the policy gradient is simply the product of a loss-dependent scalar and the gradient of the trajectory's probability.
Back-propagation can therefore replay each trajectory to compute this probability gradient in advance and apply the scalar afterwards.
As this replay also yields the trajectory's probability required by the loss, the loss computation can reuse it rather than replaying the trajectory again, merging the two trajectory replays into one.
\emph{(3) Task-level load balancing.}
A stage can further decompose into independent computation tasks, each inherently binding to its own input data rather than to the stage, which allows decoupling their GPU placement from the stage.
VAE decoding~\cite{zha2026disagfusion,huang2025ddit} within rollout, for example, which converts generated latents into video frames, operates solely on its input latents, while the loss computation and back-propagation require only the trajectory itself.
Once its inputs are ready, a task can thus run on any available GPU rather than the fixed GPUs of its stage, which makes load balancing possible at the finer granularity (i.e., tasks).

Inspired by our insights, we propose \emph{StageBreaker}, a fine-grained data-dependency-driven scheduling framework for AR video RL post-training, which schedules the true dataflow of an RL step rather than its coarse stages (Figure~\ref{fig:intro}c).
Specifically, building on asynchronous reward evaluation, we propose \emph{reward-asynchronous policy training}, which computes and back-propagates the reward-independent part of the loss without waiting for the advantages.
RL algorithms, however, clip the loss for training stability~\cite{shao2024deepseekmath}, which confines the reward-independent part to two possible forms depending on the sign of the advantage.
StageBreaker therefore retains both forms during back-propagation, which occupy negligible GPU memory under LoRA.
Once the advantage arrives, StageBreaker applies it to the matching form, yielding a policy gradient identical to that of stage-centric execution.
StageBreaker thus overlaps policy training on the actor GPUs with reward evaluation on the reward GPUs, eliminating the idle wait of the actor GPUs.
Moreover, within policy training, we design \emph{merged trajectory replay}, which further breaks the barrier between the loss computation and the back-propagation and thus replays each trajectory only once.
By exploiting the aggregation granularity of the trajectory's probability (e.g., chunk), StageBreaker feeds the probability of each chunk into the loss computation upon replaying it and then back-propagates this probability for its gradient before replaying the next chunk.
After replaying the entire trajectory, StageBreaker applies the derived loss-dependent scalar to the probability gradient, completing both the loss computation and the back-propagation within a single replay.
Rather than binding computation to GPUs by stage, our \emph{task-level placement} balances the load across the two GPU groups at the granularity of tasks.
Through a profiling-based search for the optimal placement, StageBreaker strategically moves VAE decoding and part of the policy training tasks from the overloaded actor GPUs to the underutilized reward GPUs.
Our evaluation results show that StageBreaker achieves a speedup of up to 1.41$\times$ over stage-centric execution and 1.81$\times$ over asynchronous reward evaluation.

Our main \textbf{contributions} can be summarized as follows:

\noindent $\bullet$ We characterize the challenges of video RL post-training under emerging VLM-based reward models and long videos. Even with asynchronous reward evaluation, stage-centric execution still suffers from reward-blocked policy training, redundant trajectory replay, and imbalanced GPU load.

\noindent $\bullet$ We propose StageBreaker, which schedules the true dataflow of an RL step rather than its stages. StageBreaker exploits the reward-separable policy training to overlap policy training with reward evaluation, leverages the loss-independent back-propagation to merge the two trajectory replays, and balances the load across GPUs via task-level placement.

\noindent $\bullet$ Our evaluation results show that StageBreaker achieves consistent speedups over stage-centric execution and asynchronous reward evaluation while preserving the training fidelity.

\section{Background}
\label{sec:background}

\subsection{Autoregressive Video Generation}
\label{sec:bg-video}

\noindent \textbf{Model architecture.}
Video models generate a video by iterative denoising in a latent space~\cite{wan2025wan,yang2026longlive,wu2025hunyuanvideo,teng2025magi}.
As Figure~\ref{fig:background}a shows, a text encoder embeds the prompt, a diffusion transformer (DiT) model estimates the clean latent from an initial noisy latent under this embedding, and a variational autoencoder (VAE) decodes the clean latent into video frames~\cite{peebles2023scalable,huang2025ddit}.
At the core of the video model, the DiT stacks multiple transformer blocks, each of which applies attention over the latent and the prompt embedding, followed by a feed-forward network (FFN).
With these blocks, the DiT denoises the latent from initial noise $x_{K}$ to the clean latent $x_{0}$ in $K$ steps, at each of which the DiT removes a portion of the noise from $x_{k}$ to obtain $x_{k-1}$.
Each denoising step is compute-intensive with little throughput gain from batching, as the attention and FFN of every block run over a long sequence of spatiotemporal latent tokens.
Video models therefore typically generate one video at a time, each taking a nearly constant time given the fixed number of denoising steps.

\noindent \textbf{Autoregressive generation.}
To support long-form and streaming video generation, recent works have widely explored autoregressive (AR) video models, which generate a video sequentially in $C$ chunks (i.e., a few consecutive video frames each) with the DiT-based model ~\cite{huang2025selfforcing,cui2026self,chen2026longlive,chen2025skyreels,yin2025slow,liu2026rolling}.
As Figure~\ref{fig:background}a illustrates, when generating chunk $c$, the DiT additionally attends to the key-value (KV) cache $x^{<c}$ of all preceding chunks at each denoising step, and denoises the chunk from $x^{c}_{K}$ to $x^{c}_{0}$ through the same $K$ steps.
The DiT then appends $x^{c}_{0}$ to the KV cache for the subsequent generation, and the VAE decodes it into video frames. After all $C$ chunks are generated, their frames are concatenated into the complete video.
To reduce generation cost, these models denoise each chunk in only a few steps, typically $K=4$, which brings generation to real-time frame rates~\cite{huang2025selfforcing,cui2026self,chen2026longlive}.

\subsection{RL Post-Training for Video Models}
\label{sec:bg-rl}

\begin{figure}%
  \centering
  \includegraphics[width=\linewidth]{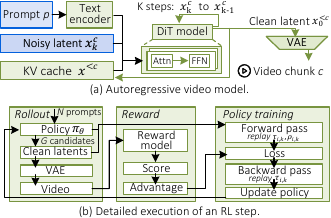}
  \vspace{-20pt}
  \caption{AR video model and its RL post-training.}
  \label{fig:background}
\end{figure}

Reinforcement learning (RL) post-training optimizes a video model for qualities such as human preference, motion quality, and prompt adherence~\cite{gao2025seedance,seedream2025seedream,lu2026raven,he2026ar,wang2026tempact,lai2026spotlight}.
RL pursues this goal over repeated steps, each of which consists of three stages.
Within a step, the current video model (i.e., policy) generates candidate videos for a batch of prompts in rollout, and the reward stage scores the qualities of candidates~\cite{xue2025dancegrpo,zhang2026astrolabe,zhang2026kvpo,wang2026leangrpo,liu2025flow}.
Policy training then updates the policy to favor candidates of higher quality.
Existing video RL systems typically execute the three stages sequentially (i.e., stage-centric execution), each over the whole batch~\cite{verl_omni,sheng2025hybridflow,dancegrpo_github,lai2026spotlight}.

Figure~\ref{fig:background}b illustrates each stage of the stage-centric execution in detail.
\emph{(1) Rollout.}
For each of $N$ prompts, the policy generates the clean latents of $G$ video candidates. The VAE then decodes them into videos for the reward stage.
\emph{(2) Reward evaluation.}
The reward model scores video candidate $i$ on the target qualities, returning the score $r_i$.
GRPO-based RL algorithms \cite{xue2025dancegrpo,zhang2026kvpo,liu2025flow} then normalize the score of each candidate within its group $\mathcal{G}(i)$, the $G$ candidates of the same prompt, into an advantage $A_i$,
\begin{equation}
A_i = \frac{r_i - \mathrm{mean}_{j \in \mathcal{G}(i)}\, r_j}{\mathrm{std}_{j \in \mathcal{G}(i)}\, r_j + \epsilon}.
\label{eq:adv}
\end{equation}
The advantage captures the relative quality of a candidate among its peers, serving as the learning signal for subsequent policy training.
\emph{(3) Policy training.}
While different RL algorithms for video models differ in their algorithmic details~\cite{dancegrpo_github,kvpo_github,wang2026leangrpo}, they follow a common GRPO-based training paradigm.
Specifically, training splits these candidates into several mini-batches and updates the weights $\theta$ of the policy $\pi_\theta$ after each.
Each update raises the probability of the trajectories (i.e., the sequences of denoising steps) that generated the candidates with positive advantages and lowers that of the others.
To this end, it minimizes a surrogate loss $\mathcal{L}(\theta)$~\cite{schulman2017proximal},
\begin{equation}
\mathcal{L}(\theta) = -\frac{1}{NGK}\sum_{i,k} \min\!\big(\rho_{i,k} A_i,\ \mathrm{clip}(\rho_{i,k}, 1-\varepsilon, 1+\varepsilon)\, A_i\big),
\label{eq:loss}
\end{equation}
where $\rho_{i,k}$ is the importance ratio of candidate $i$ at its $k$-th denoising step $\tau_{i,k}$ across all $C$ chunks.
This ratio compares the probability of $\tau_{i,k}$ under the current policy $\pi_\theta$ with that under the rollout policy $\pi_{\mathrm{old}}$,
\begin{equation}
\rho_{i,k} = \exp\!\big(\log \pi_\theta(\tau_{i,k}) - \log \pi_{\mathrm{old}}(\tau_{i,k})\big).
\label{eq:ratio}
\end{equation}
The policy generates $\tau_{i,k}$ by sampling $x^{c}_{k-1}\sim\pi_\theta(\cdot \mid x^{c}_{k}, p_i, x^{<c})$ in each chunk, and its probability $\pi_\theta(\tau_{i,k})$ therefore aggregates over all $C$ chunks as
\begin{equation}
\log \pi_\theta(\tau_{i,k}) = \sum_{c=1}^{C} \log \pi_\theta(x^{c}_{k-1} \mid x^{c}_{k}, p_i, x^{<c}).
\label{eq:logprob}
\end{equation}
As $\pi_\theta(\tau_{i,k})$ depends on all $C$ chunks, the policy obtains it only after the last chunk finishes.
Moreover, for training stability~\cite{shao2024deepseekmath,schulman2017proximal}, the clipping in Equation~\ref{eq:loss} bounds the magnitude of each update by constraining $\rho_{i,k}$ within $[1-\varepsilon, 1+\varepsilon]$, and a clipped denoising step contributes no gradient.
Back-propagating this loss $\mathcal{L}(\theta)$ yields the policy gradient and further updates the weights,
\begin{equation}
-\nabla_\theta \mathcal{L}(\theta) = \frac{1}{NGK}\sum_{i,k} A_i\, \rho_{i,k}\, \nabla_\theta \log \pi_\theta(\tau_{i,k}),
\label{eq:pg}
\end{equation}
As in conventional model training, existing systems compute the loss $\mathcal{L}(\theta)$ with a forward pass and back-propagate it with a backward pass~\cite{wang2026leangrpo,dancegrpo_github,kvpo_github}.
The forward pass replays each trajectory, re-running its denoising steps chunk by chunk under the current policy $\pi_\theta$ to derive its probability (Equation~\ref{eq:logprob}), computes the importance ratios (Equation~\ref{eq:ratio}), and then produces the loss.
As back-propagation starts only after the loss computation finishes, the forward pass cannot retain the activations of all trajectories within GPU memory until then.
The backward pass thus repeats the replay of each trajectory with gradient tracking, recomputing its activations along with its probability.
It then back-propagates the loss through the DiT with these activations to obtain the policy gradient (Equation~\ref{eq:pg}).
Note that while RL algorithms aggregate the probability at different granularities (e.g., denoising step or candidate~\cite{xue2025dancegrpo,zhang2026kvpo}), this changes only the computation of $\rho_{i,k}$, not its role as a scalar multiplier in Equation~\ref{eq:pg}.
To limit the cost of updating the weights, video RL commonly adopts low-rank adaptation (LoRA)~\cite{hu2021lora,yang2026dit,zhang2026kvpo}, which trains only a small fraction of weights.
The updated policy then serves the next RL step.

\section{Motivation}
\subsection{Preliminary Study}

Reward evaluation and policy training now account for most of each RL step latency, as two recent trends prolong both stages.

\noindent \textbf{VLM-based reward models.}
Traditional discriminative reward models assess a video in a single forward pass of a fixed scoring function~\cite{ma2025hpsv3,wu2023human,wang2026unified}, which involves little in-depth reasoning and cannot adapt its criteria to varying prompt requirements, leading to inaccurate rewards~\cite{wang2025vr}.
Recent works~\cite{wang2026unified,wang2026unifiedThink,wang2025vr,wang2026cac,he2025videoscore2} therefore turn to VLM-based reward models, which leverage a long CoT~\cite{wang2026unifiedThink,wang2026unified,bai2025qwen3} to activate the VLM's strong reasoning ability to rigorously analyze the video against the prompt requirements along multiple dimensions, thereby producing accurate and robust rewards.
Rather than judging from a fixed set of sampled frames, some of these reward models~\cite{wang2025vr} further retrieve additional frames via tool calls when examining a candidate, assessing temporal coherence across consecutive frames rather than sparse samples.
These capabilities, however, come at the cost of turning each scoring from a single forward pass into a long VLM inference, as the reward model encodes many frames of each candidate, decodes a long CoT token by token, and repeats this encoding and decoding for every frame-retrieving tool call.

\noindent\textbf{Policy training over long videos.}
AR video models extend generation from short clips of a few chunks to long videos of tens of chunks for long-form and streaming applications, and keep pushing toward minute-scale videos~\cite{cui2026self,chen2026longlive,yang2026longlive,huang2025selfforcing}.
Such long videos, however, prolong policy training with increasing video length, as its forward pass and backward pass each replay the trajectory of every candidate through all $C$$\times$$K$ denoising steps of the DiT model (\S\ref{sec:bg-rl}).
LoRA does not shorten either pass, as it reduces only the trainable weights while every denoising step still runs through the entire DiT model.

\begin{figure}[t]
  \centering
  \includegraphics[width=\linewidth]{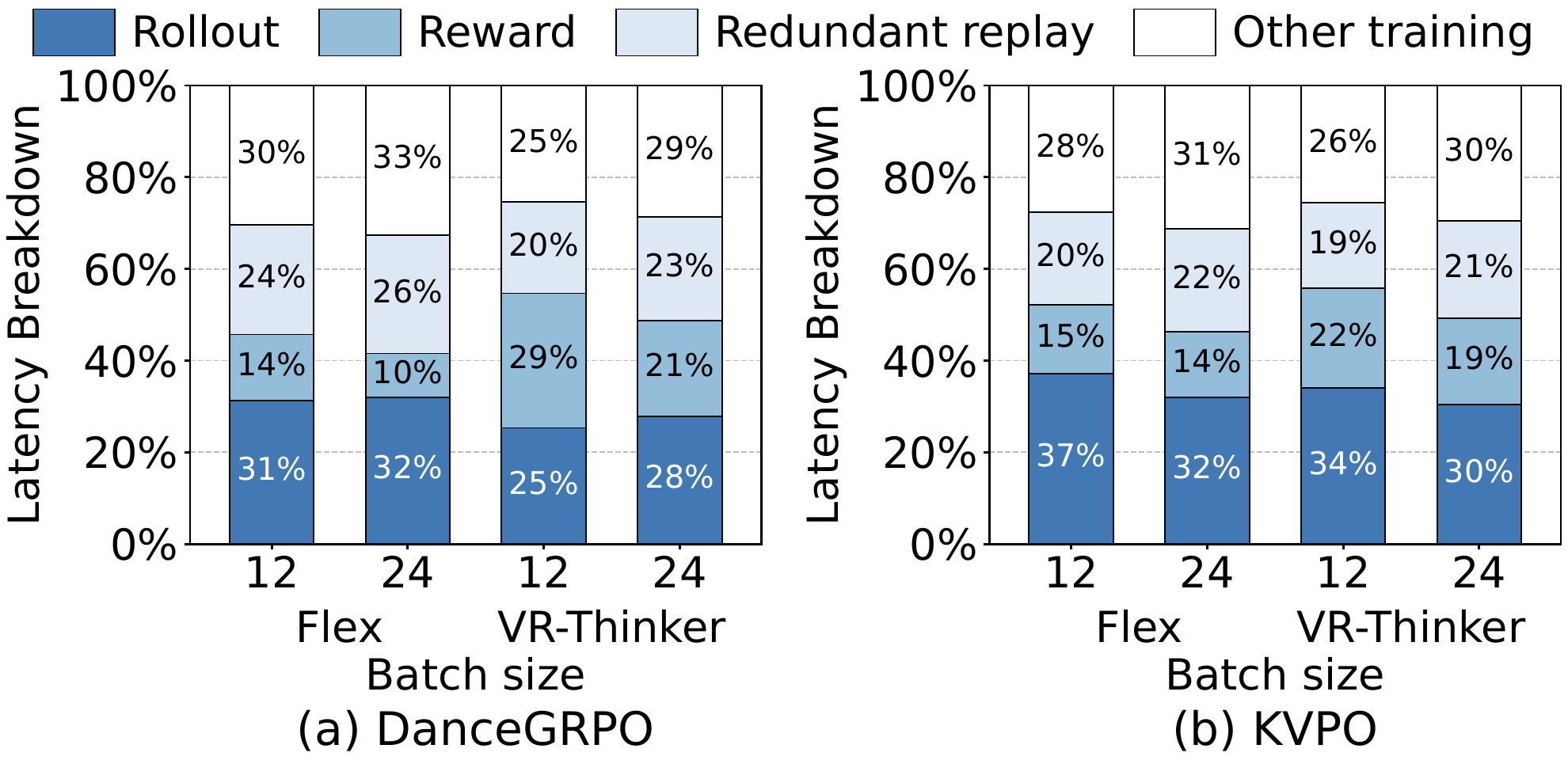}
  \vspace{-22pt}
  \caption{End-to-end RL step latency breakdown.}

  \vspace{-10pt}
  \label{fig:moti_brk}
\end{figure}

Figure~\ref{fig:moti_brk} breaks down the latency of an RL step for two video RL algorithms (DanceGRPO~\cite{xue2025dancegrpo} and KVPO~\cite{zhang2026kvpo}), each with two VLM-based reward models (UnifiedReward-Flex~\cite{wang2026unified} and VR-Thinker~\cite{wang2025vr}) under current stage-centric execution.
This execution runs reward evaluation and policy training strictly in sequence, exposing their cost entirely in the end-to-end latency of each RL step.
Across the four configurations, reward evaluation and policy training together account for 63$\sim$75\% of the RL step latency.

\subsection{Straightforward Solution and Challenges}
\label{sec:challenges}

\begin{figure}[t]
  \centering
  \includegraphics[width=\linewidth]{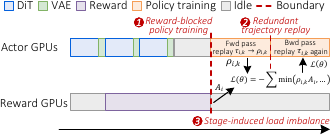}
  \vspace{-16pt}
  \caption{RL step execution in existing systems.}
  \label{fig:moti_challenge}
\end{figure}

\noindent \textbf{Asynchronous reward evaluation.}
Reward evaluation, in particular, need not lie on the critical path, as scoring a candidate depends only on its own video, which becomes available upon generation rather than after the entire rollout stage.
To take it off the critical path, a straightforward solution is asynchronous reward evaluation~\cite{verlomni_async_reward,zhu2026distrs,gao2026rollpacker} (cf. Figure~\ref{fig:moti_challenge}), which deploys the reward model on dedicated GPUs (reward GPUs) to score each candidate upon generation and leaves the remaining GPUs (actor GPUs) to run rollout and policy training.
Ideally, reward evaluation overlaps with rollout, shortening the end-to-end latency by nearly the entire reward cost. In our evaluation, however, asynchronous reward evaluation fails to deliver the expected gains and even slows RL training down by 20$\sim$29\% (cf. \S\ref{sec:evalaution}).

This solution is only an initial step away from stage-centric execution, since it removes only the stage boundary between rollout and reward evaluation.
Under the rising reward and policy training costs, the remaining stage boundaries and the stage-level GPU binding raise three challenges unaddressed by asynchronous reward evaluation.

\noindent \textbf{Reward-blocked policy training.}
Although asynchronous reward evaluation removes the stage boundary of stage-centric execution between rollout and the reward stage, it still keeps the boundary between the reward stage and policy training, making policy training wait for the reward stage to finish scoring the entire batch.
This boundary remains since policy training relies on the advantages derived from all reward scores of the entire batch (Equation~\ref{eq:loss}).
Under lightweight reward models, this boundary costs little, as scoring the batch finishes nearly with rollout.
The current long VLM inference, however, prolongs reward evaluation and keeps a substantial part of the reward cost on the critical path ahead of policy training.
The actor GPUs thus finish rollout and sit idle until the last score of the batch arrives, which takes 13\% of each RL step in our measurement and leaves 36\% of the reward cost on the critical path (\S\ref{sec:eval-analysis}).

\noindent\textbf{Redundant trajectory replay.}
Within policy training, existing RL systems keep a further boundary between the loss computation and back-propagation, which results in redundant trajectory replay (\S\ref{sec:bg-rl}).
Inheriting this boundary from conventional model training, these systems start back-propagation only after computing the complete loss, and thus take two passes over every trajectory.
The forward pass first replays the trajectory to obtain the probabilities $\pi_\theta(\tau_{i,k})$ for the loss computation.
As limited GPU memory cannot retain the activations of every trajectory until the loss computation finishes (\S\ref{sec:bg-rl}), the backward pass has to repeat the entire replay to rebuild the activations and then derive the gradient from them.
As long videos prolong every replay, the resulting redundant trajectory replay accounts for 19$\sim$26\% of the RL step latency (Figure~\ref{fig:moti_brk}), a substantial waste of GPU computation.

\noindent\textbf{Stage-induced load imbalance.}
Each stage binds its computation tasks to a fixed set of GPUs regardless of the load across GPUs, resulting in severe load imbalance and GPU underutilization.
Asynchronous reward evaluation, for example, disaggregates the reward stage onto dedicated GPUs due to the rising reward cost but follows existing RL systems and still allocates GPUs at stage granularity, confining these reward GPUs to scoring alone.
Moreover, candidates arrive at these reward GPUs in a stream as rollout generates them, keeping each scoring batch small, and the memory-bound CoT decoding of the VLM thus underutilizes these GPUs even during scoring.
The long videos, in turn, prolong every task carried by the actor GPUs, from DiT and VAE decoding in rollout to policy training.
Once scoring finishes, the reward GPUs therefore sit idle throughout the prolonged policy training, while the actor GPUs remain overloaded.
Under a fixed GPU budget, the reward GPUs stay busy for only 36\% of each RL step (cf. \S\ref{sec:eval-analysis}), and this idle GPU time makes asynchronous reward evaluation 22\% slower than the colocated baseline.

\begin{figure}[t]
  \centering
  \includegraphics[width=\linewidth]{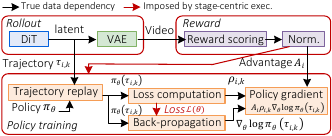}
  \vspace{-16pt}
  \caption{Dataflow of an RL step.}
  \label{fig:dataflow}
\end{figure}

\subsection{Key Insights}
\label{sec:insights}
The three challenges all stem from stage-centric execution, whose boundaries and GPU binding impose constraints coarser than the true data dependencies within an RL step.
In practice, while most of the computation after a boundary can proceed without the awaited result, the boundary still stalls all of it.
As Figure~\ref{fig:dataflow} shows, in policy training, for example, the boundaries stall the entire policy gradient computation due to its dependence on the advantage and the importance ratio.
This dependence, however, involves only the final multiplication, which reveals the separability of the two from the rest of the computation.
Similarly, each computation task depends only on its inputs, which suggests a placement beyond the GPUs of its stage.
Our key insight is therefore to schedule the true dataflow, not the stages, shortening the critical path and filling the idle GPUs.

\noindent \textbf{Reward-separable policy training.}
As Equation~\ref{eq:loss} and Figure \ref{fig:dataflow} show, the surrogate loss comprises two factors, of which only the advantage $A_i$ depends on reward scores.
The other factor, the importance ratio $\rho_{i,k}$, relies solely on the trajectory and the policy.
Given their distinct dependencies, the loss computation is separable into these two factors.
Policy training can therefore compute and back-propagate the importance ratio in advance, and apply the advantage through a single scalar multiplication once it arrives.
It thereby overlaps with reward evaluation, which removes the idle wait of the actor GPUs and takes the reward cost entirely off the critical path.

\noindent \textbf{Loss-independent back-propagation.}
Equation~\ref{eq:pg} and Figure \ref{fig:dataflow} illustrate that the policy gradient computation is likewise separable, as the loss computation contributes only $\rho_{i,k}$, a scalar multiplier of the gradient $\nabla_\theta \log \pi_\theta(\tau_{i,k})$.
This gradient itself depends on nothing but the trajectory and the policy, and back-propagation can thus compute it in advance and apply $\rho_{i,k}$ afterwards, rather than waiting for the loss.
This gradient requires replaying each trajectory, which also yields the probabilities $\log \pi_\theta(\tau_{i,k})$ for loss.
The loss computation can therefore reuse these probabilities rather than recompute them, merging the two trajectory replays. %

\noindent\textbf{Task-level load balancing.}
A stage further decomposes into independent computation tasks, each depending only on its own inputs rather than on the entire stage, which makes the stage-level GPU binding coarser than necessary.
As Figure~\ref{fig:dataflow} shows, VAE decoding, for example, depends only on the clean latents of a candidate, but the rollout stage binds it to the actor GPUs while the underutilized reward GPUs could run it instead.
As VAE decoding takes about 30\% of the rollout time (\S\ref{sec:eval-analysis}), moving it alone would shift a substantial load from the actor GPUs to the reward GPUs.
Moreover, reward scoring starts once these VAE decodings finish and thus runs on a large batch, raising the throughput of the memory-bound CoT decoding without delaying the completion of reward evaluation.
Similarly, the loss computation and back-propagation of a trajectory depend only on the trajectory itself and the policy, but still stay on the overloaded actor GPUs.
Such tasks can thus strategically move from the overloaded actor GPUs to the idle reward GPUs, mitigating GPU underutilization.

\section{Designs}

\begin{figure}%
  \centering
  \includegraphics[width=0.99\linewidth]{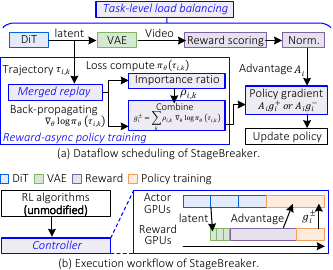}
  \vspace{-7pt}
  \caption{Overview of StageBreaker.}
  \vspace{-10pt}
  \label{fig:overview}
\end{figure}

\subsection{Overview}
\label{sec:overview}
Inspired by the above analysis, we propose StageBreaker (cf. Figure \ref{fig:overview}a), a data-dependency-driven scheduling framework for AR video RL post-training.
StageBreaker executes each RL step as a dataflow of computation tasks along their true data dependencies rather than stage boundaries, and dispatches these tasks across GPUs at task level rather than stage granularity.

Figure~\ref{fig:overview}a shows the dataflow of an RL step under StageBreaker.
Following reward-separable policy training, StageBreaker designs \emph{reward-asynchronous policy training} (\S\ref{sec:async-train}), which computes and back-propagates the reward-independent part of each loss term (i.e., the importance ratio $\rho_{i,k}$) on the actor GPUs without waiting for the advantage $A_i$ from the reward GPUs.
Once the advantage arrives, StageBreaker scales the resulting gradient by it, yielding the policy gradient (Equation~\ref{eq:pg}) identical to that of stage-centric execution.
Under the clipped form (Equation~\ref{eq:loss}), however, this gradient varies with the sign of the advantage.
StageBreaker thus retains it for either sign, $g^{+}_i$ and $g^{-}_i$, applying the advantage to the matching one.
Moreover, within the reward-independent part, \emph{merged trajectory replay} (\S\ref{sec:merged-replay}) exploits loss-independent back-propagation to break the barrier between the loss computation and back-propagation.
Back-propagation replays each trajectory with gradient tracking to compute the gradient of its probability $\nabla_\theta \log \pi_\theta(\tau_{i,k})$.
The loss computation then reuses the probability $\pi_\theta(\tau_{i,k})$ produced by this same replay for the importance ratio $\rho_{i,k}$, rather than replaying the trajectory again.
StageBreaker finally combines the importance ratio $\rho_{i,k}$ and $\nabla_\theta \log \pi_\theta(\tau_{i,k})$ at every denoising step (Equation~\ref{eq:branch}).
Rather than binding stages to GPUs, \emph{task-level load balancing} (\S\ref{sec:placement}) balances the load across the two GPU groups at the granularity of tasks.
Beyond asynchronous reward evaluation, StageBreaker equips the reward GPUs with a VAE decoder and a training engine, enabling them to take over VAE decoding and the loss computation and back-propagation of trajectories from the overloaded actor GPUs.
A profiling-based search then determines the split of these tasks between the two GPU groups, adopting the placement minimizing the end-to-end step latency.

\noindent\textbf{Workflow.}
Figure~\ref{fig:overview}b shows RL step execution of StageBreaker, where a controller above the engines orchestrates the execution of all tasks and keeps the RL algorithm on top unmodified.
As the rollout engine on the actor GPUs generates the latent of each candidate, the controller dispatches its latents to the reward GPUs, where the VAE decoder converts them into the video and the reward engine then scores it, while rollout continues.
When rollout completes, the actor GPUs start policy training immediately, running the loss computation and back-propagation of each trajectory in a single replay to obtain the gradient of its reward-independent part without waiting for reward scores.
As reward scores arrive, the controller applies the advantages to the corresponding candidates, and the policy updates once the policy gradient of a mini-batch becomes available.
After scoring, the reward GPUs likewise take part in remaining policy training.
StageBreaker thereby only reschedules the execution of the RL step, preserving the training semantics of stage-centric execution exactly.

\subsection{Reward-Asynchronous Policy Training}

\label{sec:async-train}

\begin{figure}%
  \centering
  \includegraphics[width=0.99\linewidth]{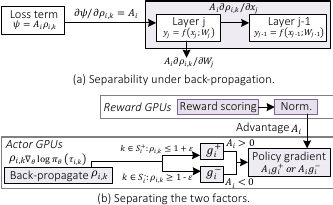}
  \vspace{-6pt}
  \caption{Reward-asynchronous policy training.}
  \label{fig:async}
\end{figure}

StageBreaker separates each loss term into its reward-dependent and reward-independent parts, which the reward GPUs and the actor GPUs compute respectively.
We first prove that the gradient of each LoRA weight still preserves this separability, then describe the separated execution on GPUs, and finally the overlap of policy training with reward evaluation.

\noindent\textbf{Separability under back-propagation.}
Ahead of the advantage, StageBreaker back-propagates the importance ratio $\rho_{i,k}$ of each term rather than the complete term $A_i\rho_{i,k}$ itself.
This back-propagation passes through every layer of the DiT, linear and nonlinear, before reaching each LoRA weight.
We now prove that the gradient of each LoRA weight still preserves this separability, with $A_i$ remaining its scalar factor.
As Figure \ref{fig:async}a illustrates, for an unclipped term $\psi = A_i\rho_{i,k}$ of the loss, back-propagation starts from the derivative $\partial\psi/\partial\rho_{i,k} = A_i$, a scalar, and passes it down through the layers of the DiT.
Let layer $j$ compute output activations $y_j = f_j(x_j, W_j)$ from input activations $x_j$ and weights $W_j$, where $f_j$ can be linear or nonlinear.
At every layer, back-propagation applies the chain rule
\begin{equation}
\frac{\partial \psi}{\partial x} = \sum_{j} \Big(\frac{\partial y_j}{\partial x}\Big)^{\!\top} \frac{\partial \psi}{\partial y_j},
\quad
\frac{\partial \psi}{\partial W_j} = \Big(\frac{\partial y_j}{\partial W_j}\Big)^{\!\top} \frac{\partial \psi}{\partial y_j},
\label{eq:vjp}
\end{equation}
where $j$ ranges over the layers taking $x$ as input.
Equation~\ref{eq:vjp} implies that $A_i$ remains a scalar factor across each layer.
To be specific, the input $x_j$ and weights $W_j$ of the layer determine both Jacobians, $\partial y_j/\partial x$ and $\partial y_j/\partial W_j$ (e.g., $W_j$ and $x_j$ for a linear layer $y_j = W_j x_j$).
As $A_i$ appears in neither $x_j$ nor $W_j$, both Jacobians are independent of $A_i$.
Equation~\ref{eq:vjp} is therefore linear in $\partial\psi/\partial y_j$, which preserves any scalar factor of $\partial\psi/\partial y_j$.
Hence, for $\partial\psi/\partial y_j = A_i h_j$, Equation~\ref{eq:vjp} yields $\partial\psi/\partial x = A_i \sum_j (\partial y_j/\partial x)^{\top} h_j$ and $\partial\psi/\partial W_j = A_i\,(\partial y_j/\partial W_j)^{\top} h_j$, both keeping $A_i$ as a scalar factor.
Appendix~\ref{app:backward} verifies both properties for the concrete DiT layers, including attention.
Therefore, starting from $\partial\psi/\partial\rho_{i,k} = A_i$, the back-propagation keeps $A_i$ as a scalar factor through every layer and arrives at $\partial\psi/\partial W_j = A_i\,\partial\rho_{i,k}/\partial W_j$ for each LoRA weight.
Moreover, Equation~\ref{eq:pg} merely sums these per-term gradients over all $i$ and $k$.
Back-propagating each $\rho_{i,k}$ alone to obtain $\nabla_\theta\rho_{i,k}$, scaling it by $A_i$ afterwards, and summing the results thus recovers the exact policy gradient.
Note that the clipping of Equation~\ref{eq:loss} does not break this separability, since it only zeroes the gradient of a clipped term and keeps $A_i$ a scalar factor.

\noindent\textbf{Separating the two factors.}
With the separability proven, StageBreaker computes the gradient of the importance ratio and the advantage separately and combines them afterwards.
However, the clipping in Equation~\ref{eq:loss} complicates this separation, as the clipping result depends on the advantage $A_i$, while back-propagation runs independently of $A_i$.
StageBreaker therefore prepares each clipping branch in advance.
Specifically, the clipping branch depends on $\rho_{i,k}$ and the sign of $A_i$, contributing $\rho_{i,k}\,\nabla_\theta \log \pi_\theta(\tau_{i,k})$ if $\rho_{i,k} \le 1+\varepsilon$ under a positive advantage, or if $\rho_{i,k} \ge 1-\varepsilon$ under a negative one, and nothing otherwise.
StageBreaker thus accumulates two gradients for each candidate~$i$, $g^{+}_i$ for a positive advantage and $g^{-}_i$ for a negative one,
\begin{equation}
g^{\pm}_i = \sum_{k \in S^{\pm}_i} \rho_{i,k}\,\nabla_\theta \log \pi_\theta(\tau_{i,k}),
\label{eq:branch}
\end{equation}
where $g^{+}_i$ sums over $S^{+}_i = \{k \mid \rho_{i,k} \le 1+\varepsilon\}$, the steps contributing under a positive advantage, and $g^{-}_i$ over $S^{-}_i = \{k \mid \rho_{i,k} \ge 1-\varepsilon\}$, those under a negative one.
Holding $g^{\pm}_i$ on GPUs is affordable, as each gradient spans only the trainable LoRA weights and thus occupies negligible memory.
Once the advantages arrive, StageBreaker takes $A_i g^{+}_i$ if $A_i > 0$ and $A_i g^{-}_i$ otherwise for each candidate, and accumulates the result over the mini-batch following Equation~\ref{eq:pg}.
This combination costs only a scalar multiplication per candidate, negligible against back-propagation.

\noindent\textbf{Overlapping training with reward stage.}
As Figure~\ref{fig:async}b shows, the actor GPUs compute and back-propagate the importance ratios of each mini-batch while the reward GPUs score its groups, and the two GPU groups synchronize only at the optimizer step.
StageBreaker thereby removes the stage-level reward barrier and the idle wait it imposes on the actor GPUs.

\begin{figure}%
  \centering
  \includegraphics[width=0.99\linewidth]{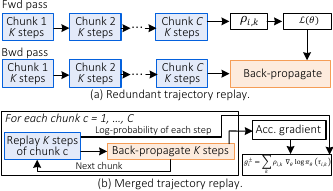}
  \vspace{-6pt}
  \caption{Trajectory replay in existing systems and StageBreaker.}
  \vspace{-10pt}
  \label{fig:merged}
\end{figure}

\subsection{Merged Trajectory Replay}
\label{sec:merged-replay}

StageBreaker further eliminates the redundant trajectory replay within policy training (\S\ref{sec:challenges}).
As Figure~\ref{fig:merged}a illustrates, existing RL systems replay each trajectory twice, first without gradient tracking to obtain the probabilities $\pi_\theta(\tau_{i,k})$ for the importance ratios $\rho_{i,k}$ and the loss $\mathcal{L}(\theta)$, and then with gradient tracking to back-propagate $\mathcal{L}(\theta)$.
The two replays arise as these systems serialize the loss computation and back-propagation, back-propagating only the fully computed loss.

Back-propagation, however, need not wait for the complete loss.
Building on the separated advantage, StageBreaker further separates the gradient of the reward-independent part, $\nabla_\theta \rho_{i,k} = \rho_{i,k}\,\nabla_\theta \log \pi_\theta(\tau_{i,k})$ (Equation~\ref{eq:branch}), in which $\rho_{i,k}$ is also only a scalar multiplier.
Back-propagation therefore replays each trajectory directly to compute $\nabla_\theta \log \pi_\theta(\tau_{i,k})$, without waiting for $\rho_{i,k}$ or the loss.
This replay inherently produces the probabilities $\pi_\theta(\tau_{i,k})$, which can be reused for the importance ratios.
StageBreaker thus merges the two trajectory replays into a single replay.

\noindent\textbf{Replay procedure.}
As the activations of all trajectories rarely fit within GPU memory (\S\ref{sec:bg-rl}), the merged replay proceeds at chunk granularity (Figure~\ref{fig:merged}b), releasing the activations of each chunk before replaying the next.
Within each chunk, it back-propagates each denoising step $\tau^{c}_{i,k}$ upon replaying it, buffering its gradient $\nabla_\theta \log \pi_\theta(\tau^{c}_{i,k})$.
For chunk $c$, it re-runs the $K$ denoising steps under the current policy, accumulating the log-probability $\log \pi_\theta(\tau^{c}_{i,k})$ of each step $k$ into $\log \pi_\theta(\tau_{i,k})$ based on Equation~\ref{eq:logprob}.
It then back-propagates each step, likewise accumulating the gradient $\nabla_\theta \log \pi_\theta(\tau^{c}_{i,k})$ into $\nabla_\theta \log \pi_\theta(\tau_{i,k})$ in a buffer $g_{i,k}$.
After replaying the entire video, StageBreaker computes $\rho_{i,k}$ from $\log \pi_\theta(\tau_{i,k})$ (Equation~\ref{eq:ratio}) and adds $\rho_{i,k}\,g_{i,k}$ to $g^{\pm}_i$ according to $S^{\pm}_i$ (Equation~\ref{eq:branch}).
For RL algorithms aggregating the probability at other granularities (\S\ref{sec:bg-rl}), StageBreaker analogously accumulates the gradient of each aggregated unit and scales it by its importance ratio once the aggregation completes.

The merged replay thus costs no more computation than the second replay of existing RL systems (Figure~\ref{fig:merged}a), and adds only a small GPU memory overhead, the $K$ per-step buffers $g_{i,k}$ of the current trajectory, each the size of the LoRA weights.

\subsection{Task-Level Load Balancing}

\label{sec:placement}

StageBreaker refines the current GPU binding from stage granularity to task granularity, placing each task according to the load of the two GPU groups rather than its stage.
Following this principle, StageBreaker moves two kinds of tasks from the overloaded actor GPUs to the underutilized reward GPUs, VAE decoding in rollout and the loss computation and back-propagation in policy training (policy training tasks).

As VAE decoding depends only on the clean latents of a candidate, StageBreaker transfers these small latents to the reward GPUs and decodes them there before scoring.
The reward GPUs then score the decoded videos as one batch rather than individually upon arrival, which accelerates the memory-bound CoT decoding and thus does not delay reward evaluation.
Similarly, the policy training tasks of a trajectory (\S\ref{sec:merged-replay}) depend only on the trajectory itself and the current policy.
StageBreaker therefore assigns policy training tasks to the reward GPUs once they finish scoring, and transfers the resulting $g^{\pm}_i$ back to the actor GPUs.
Moreover, the loss computation of some RL algorithms involves additional forward passes under frozen policies (e.g., the replay under the reference policy for the KL term~\cite{zhang2026kvpo,liu2025flow,shao2024deepseekmath}), which StageBreaker likewise moves to the reward GPUs.

\noindent\textbf{Scheduling.}
To minimize the RL step latency, StageBreaker must determine the split of VAE decodings and policy training tasks between the two GPU groups.
To this end, StageBreaker builds a profiling-based simulator to search for the optimal placement.
StageBreaker profiles the execution time of every computation task on the target setup, and the simulator executes the dataflow of an RL step under a given split to estimate the end-to-end step latency.
Owing to the small search space and the fast simulation \cite{zhu2026distrs,zhong2024distserve}, StageBreaker enumerates all possible placements and adopts the one with the best performance.
Our controller then dispatches tasks by the derived placement at runtime.

\subsection{Implementation}
We implement StageBreaker on top of verl-omni~\cite{verl_omni}, a widely used RL post-training framework.
We reuse the rollout, reward, and training engines of verl-omni for all computation, while implementing our controller above these engines to schedule computation tasks.
For task-level load balancing (\S\ref{sec:placement}), we further equip the reward GPUs with a VAE decoder and a lightweight replica of the training engine, holding only the DiT and a LoRA copy and receiving the updated LoRA weights after each optimizer step.
Moreover, StageBreaker integrates two RL algorithms, DanceGRPO and KVPO, without modifying either of them. StageBreaker also readily extends to other AR video RL algorithms, as it reschedules only the execution of the RL step rather than the algorithm itself.

\section{Evaluation}
\label{sec:evalaution}

\subsection{Experiment setup}

\label{sec:eval-setup}

\noindent\textbf{Testbed.}
We conduct all experiments with NVIDIA RTX PRO 5000 Blackwell GPUs (72\,GB GDDR7 each) \cite{nvidia2025rtxpro5000}, interconnected via PCIe~5.0~$\times$16.
The host server has dual-socket 5th-generation AMD EPYC CPUs (384 cores in total) and 2.4\,TB of DDR5 memory.
Unless otherwise specified, each experiment employs six GPUs.

\noindent\textbf{Compared systems.}
We compare StageBreaker against the two popular execution schemes.
\texttt{(1) Colocated} follows stage-centric execution, the default of verl-omni, running rollout \cite{verl_omni,sheng2025hybridflow}, reward evaluation, and policy training sequentially over the entire batch on all GPUs.
\texttt{(2) AsyncReward} enables the asynchronous reward evaluation \cite{verlomni_async_reward}, which dedicates two GPUs to the reward model (reward GPUs) and the remaining four to rollout and policy training (actor GPUs).
To understand the contribution of each design, we further break StageBreaker down by enabling its designs progressively.
\texttt{(3) AsyncTrain} follows the GPU allocation of \texttt{AsyncReward} and adopts reward-asynchronous policy training (\S\ref{sec:async-train}), \texttt{(4) MergedReplay} further integrates merged trajectory replay (\S\ref{sec:merged-replay}), and \texttt{(5) StageBreaker} completes all three designs with task-level load balancing (\S\ref{sec:placement}).
All systems run the same engines with the same parallelism strategy, LoRA configuration, and hyperparameters, and thus differ only in the execution of the RL step.

\noindent\textbf{Workloads.}
We evaluate two RL algorithms, DanceGRPO~\cite{xue2025dancegrpo} and KVPO~\cite{zhang2026kvpo}, each with two VLM-based reward models, UnifiedReward-Flex (UR-Flex)~\cite{wang2026unified} and VR-Thinker (VRT)~\cite{wang2025vr}.
We employ Wan2.1-1.3B Self-Forcing~\cite{huang2025selfforcing} for DanceGRPO and LongLive-1.3B~\cite{yang2026longlive} for KVPO, both few-step AR video models trained with LoRA.
We take the prompts for RL training from the official prompt sets of the two reward models, VideoGen-RewardBench~\cite{videogen_rewardbench,liu2025flow,liu2025improving} for VR-Thinker and Pref-GRPO~\cite{pref_grpo,Pref-GRPO&UniGenBench} for UnifiedReward-Flex.
By default, each candidate video has 21 latent frames (81 pixel frames) for DanceGRPO and 33 latent frames (129 pixel frames) for KVPO at 480p, and each RL step processes a batch of 12 candidate videos, split into two mini-batches for updating the policy \cite{xue2025dancegrpo,zhang2026kvpo,dancegrpo_github,kvpo_github}.
We keep the remaining hyperparameters at the defaults of each algorithm.
We further vary the video length, resolution, batch size, and number of mini-batches to examine the compared systems under different settings.
We report the end-to-end latency of an RL step averaged over eight steps after a warm-up step.

\subsection{Overall Performance}

\begin{figure}[t]
  \centering
  \includegraphics[width=0.99\linewidth]{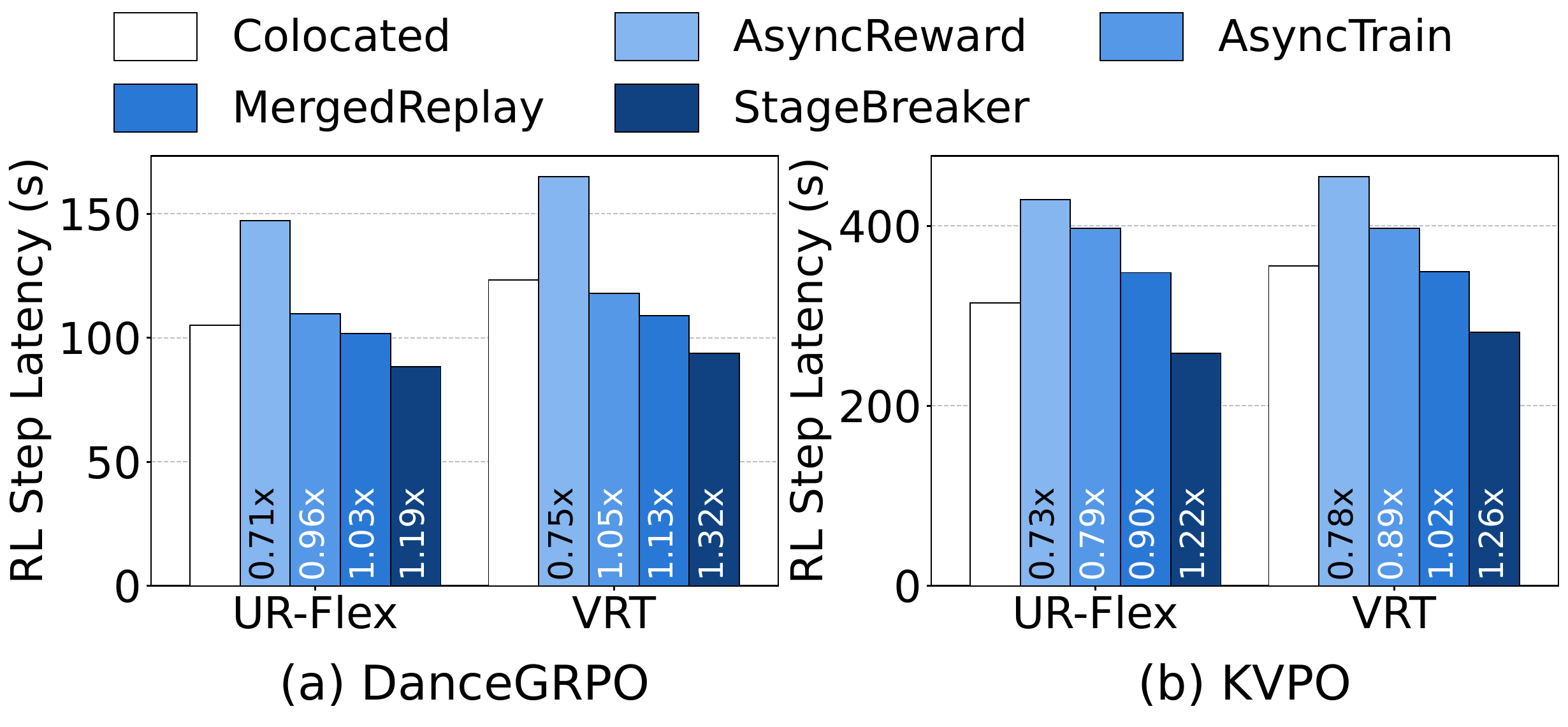}
  \vspace{-10pt}
  \caption{Performance across RL algorithms and reward models.}
  \vspace{-16pt}
  \label{fig:eval_overall}
\end{figure}

Figure~\ref{fig:eval_overall} compares the end-to-end latency of an RL step under the default configuration across the two RL algorithms and the two reward models.
Although \texttt{AsyncReward} attempts to take the reward cost off the critical path by disaggregating the reward stage onto dedicated GPUs, it leaves multiple challenges unaddressed and thus runs even slower than \texttt{Colocated} under the same GPU budget, reaching only 0.71$\sim$0.78$\times$ its performance.
The reward GPUs sit idle for most of the step, while reward-blocked policy training also leaves the actor GPUs idle, as our following analysis shows in \S\ref{sec:eval-analysis}.
\texttt{AsyncTrain} addresses reward-blocked policy training, which computes the reward-independent part before the reward scores arrive and thereby overlaps policy training with reward evaluation.
It thus achieves a speedup of up to 1.40$\times$ over \texttt{AsyncReward} and up to 1.05$\times$ over \texttt{Colocated}.
VR-Thinker brings greater performance benefits, as its longer CoT prolongs the scoring of each candidate and thereby hides more of the reward cost behind policy training.
However, \texttt{AsyncTrain} still lags behind \texttt{Colocated} in most cases, reaching only 0.79$\sim$0.96$\times$ its performance, as the barrier between the loss computation and the back-propagation still forces the actor GPUs to replay every trajectory twice.
\texttt{MergedReplay} breaks this barrier and eliminates these costs with merged trajectory replay.
It therefore achieves a speedup of 1.08$\sim$1.14$\times$ over \texttt{AsyncTrain}, and reaches 0.90$\sim$1.13$\times$ the performance of \texttt{Colocated}.
\texttt{StageBreaker} further addresses stage-induced load imbalance with task-level load balancing, which moves a portion of VAE decoding and policy training tasks to the idle reward GPUs, thereby improving utilization across all GPUs.
With all three designs integrated, \texttt{StageBreaker} achieves a speedup of 1.19$\sim$1.32$\times$ over \texttt{Colocated} and 1.62$\sim$1.76$\times$ over \texttt{AsyncReward}.

\subsection{Sensitivity Analysis}
\label{sec:eval-sensitivity}

\begin{figure}[t]
  \centering
  \includegraphics[width=0.99\linewidth]{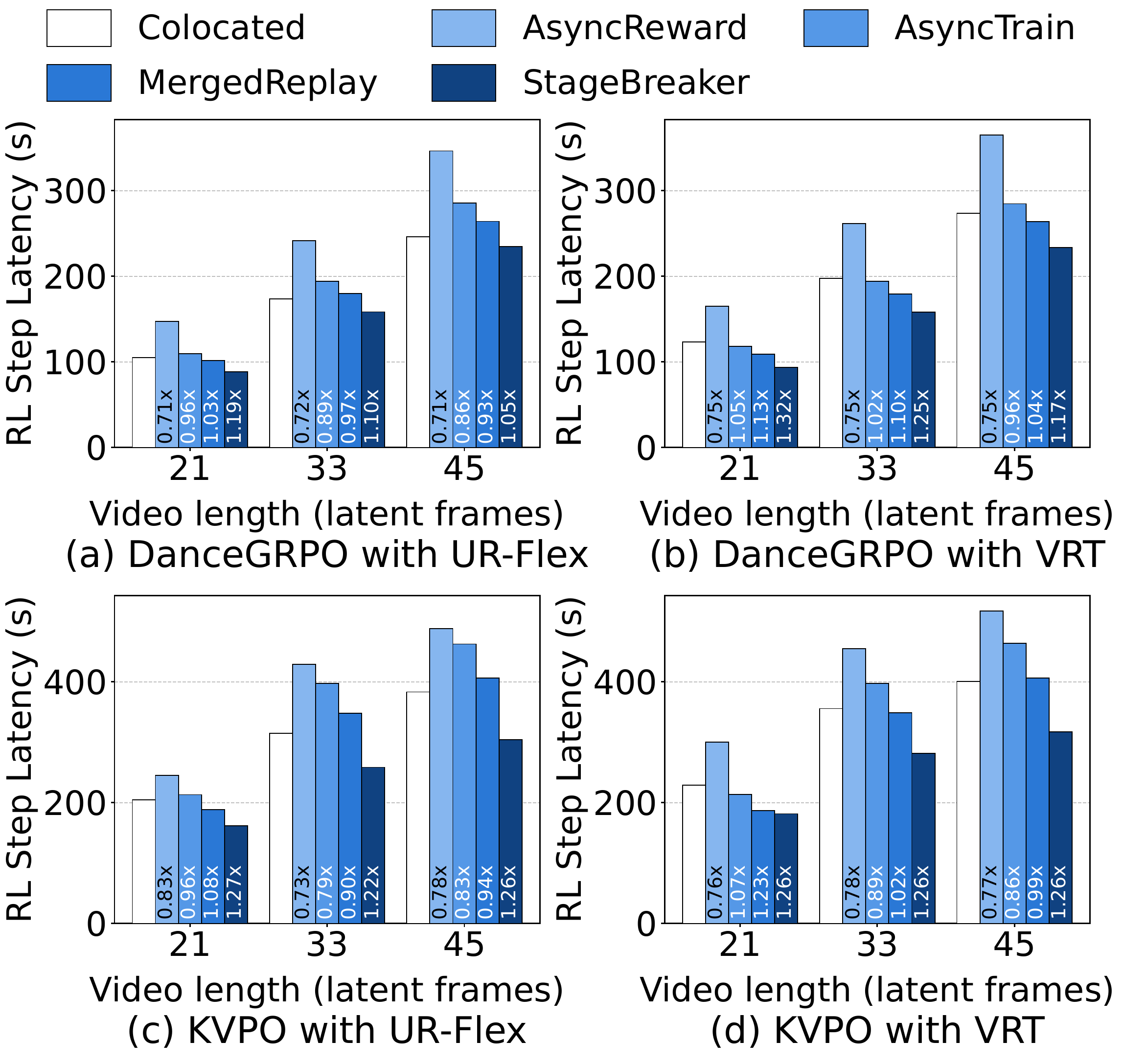}
  \vspace{-10pt}
  \caption{RL step latency with varying video lengths.}
  \vspace{-14pt}
  \label{fig:eval_video_length}
\end{figure}

\noindent \textbf{Impact of video length.}
Figure~\ref{fig:eval_video_length} varies the video length from 21 to 45 latent frames (81 to 177 pixel frames), which covers the typical lengths of mainstream video models \cite{wu2025hunyuanvideo,wan2025wan,yang2026longlive}.
Longer videos raise the cost of policy training, as each trajectory replay runs more denoising steps through the DiT.
Under \texttt{AsyncReward}, the actor GPUs pay this growing cost twice due to redundant trajectory replay, while the reward GPUs sit idle throughout the longer policy training.
\texttt{StageBreaker} targets exactly this growing cost with merged trajectory replay and task-level load balancing.
\texttt{StageBreaker} therefore consistently outperforms both baselines across all lengths, achieving a speedup of 1.05$\sim$1.32$\times$ over \texttt{Colocated} and 1.48$\sim$1.76$\times$ over \texttt{AsyncReward}.
Among the three designs, \texttt{AsyncTrain} contributes 1.06$\sim$1.41$\times$ over \texttt{AsyncReward}, more with VR-Thinker whose scoring takes longer.
\texttt{MergedReplay} adds a steady 1.08$\sim$1.14$\times$ over \texttt{AsyncTrain} at all lengths.
\texttt{StageBreaker} brings another 1.03$\sim$1.35$\times$ over \texttt{MergedReplay} by moving the longer policy training tasks onto the underutilized reward GPUs.
Across the two reward models, \texttt{StageBreaker} sustains a speedup of 1.17$\sim$1.32$\times$ with VR-Thinker and 1.05$\sim$1.27$\times$ with UnifiedReward-Flex over the three lengths.

\begin{figure}[t]
  \centering
  \includegraphics[width=0.99\linewidth]{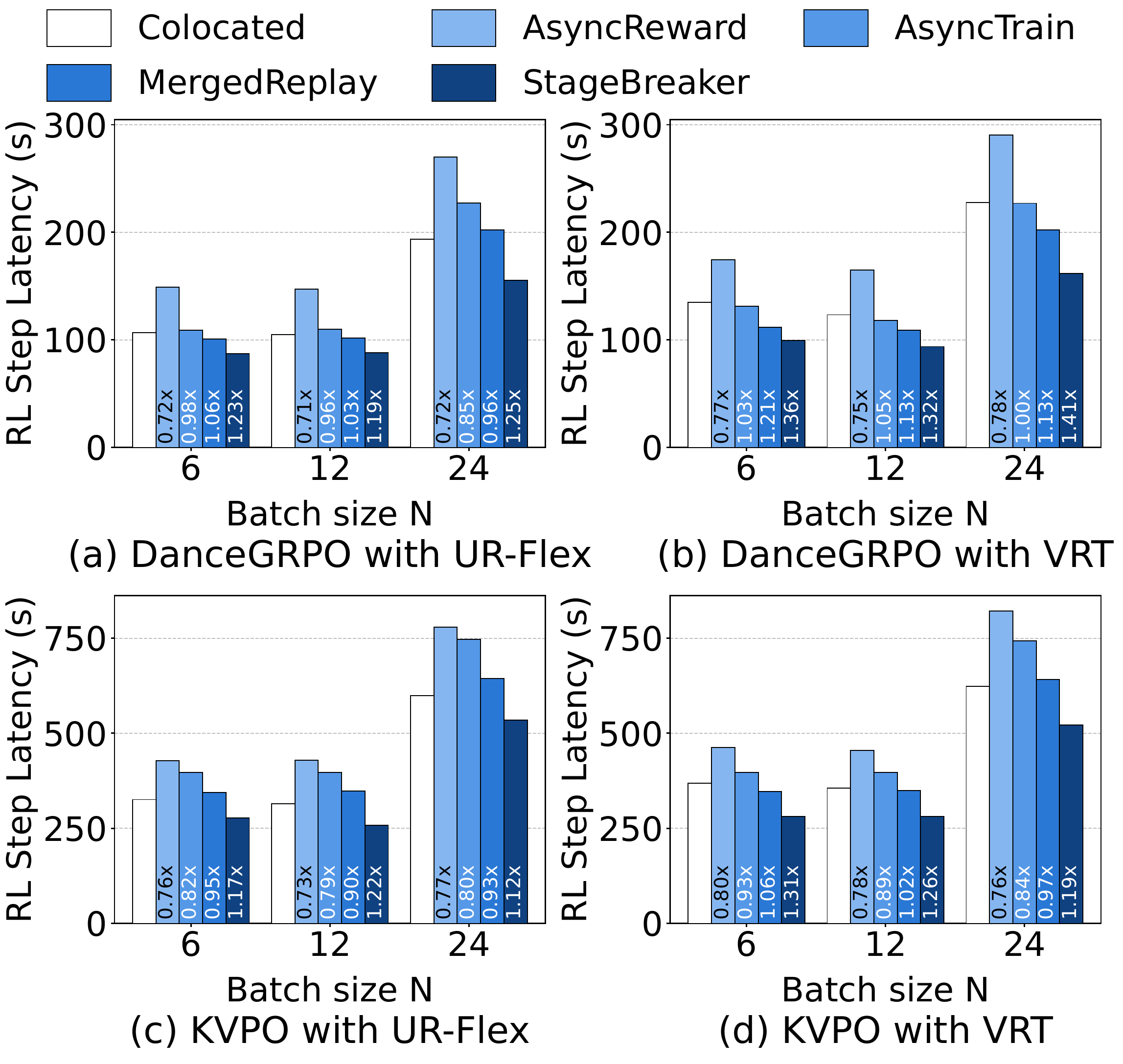}
  \vspace{-10pt}
  \caption{RL step latency with varying batch sizes.}
  \vspace{-12pt}
  \label{fig:eval_batch_size}
\end{figure}

\begin{figure}[t]
  \centering
  \includegraphics[width=0.99\linewidth]{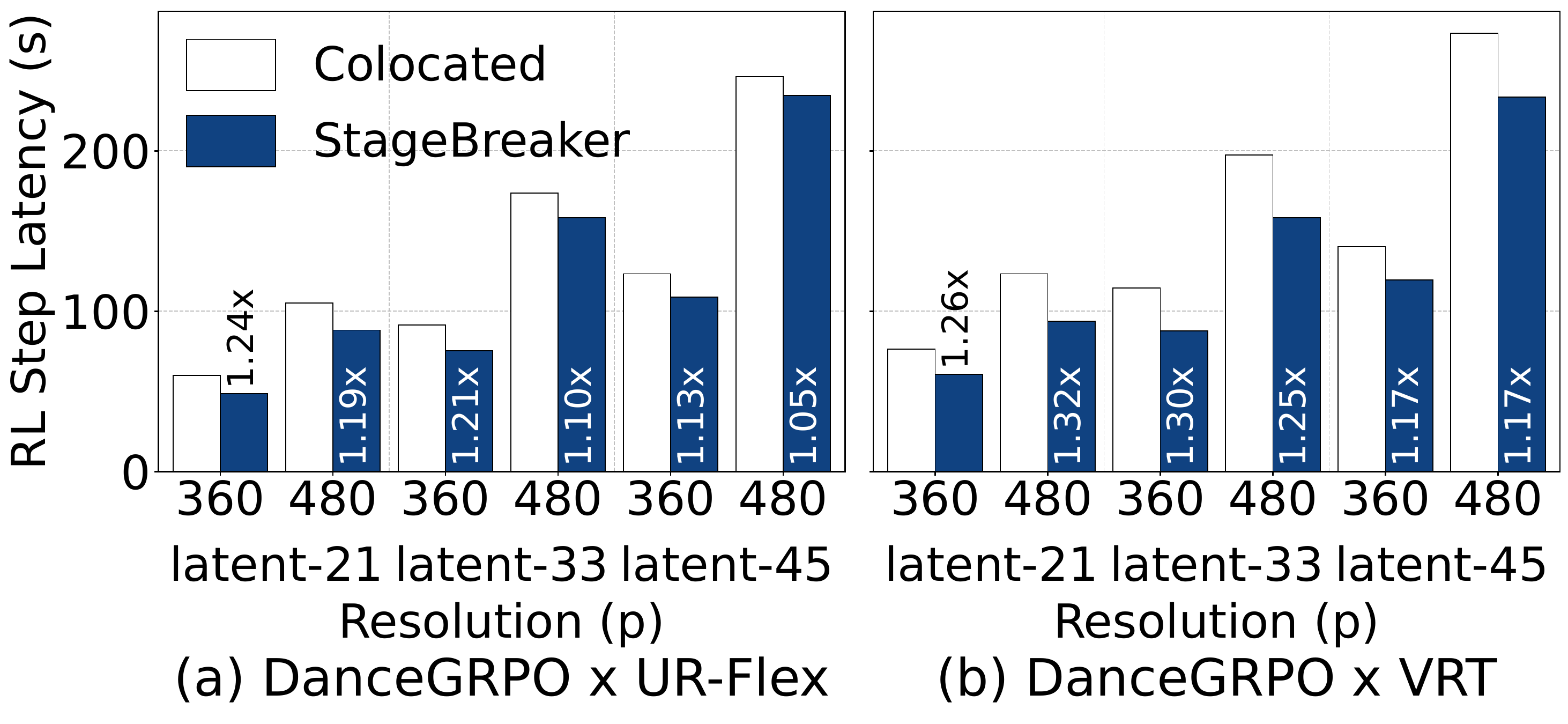}
  \vspace{-10pt}
  \caption{RL step latency with varying video resolutions.}
  \vspace{-12pt}
  \label{fig:eval_resolution}
\end{figure}

\begin{figure}[t]
  \centering
  \begin{minipage}[t]{0.49\linewidth}
    \vspace{0pt}
    \centering
    \includegraphics[width=\linewidth]{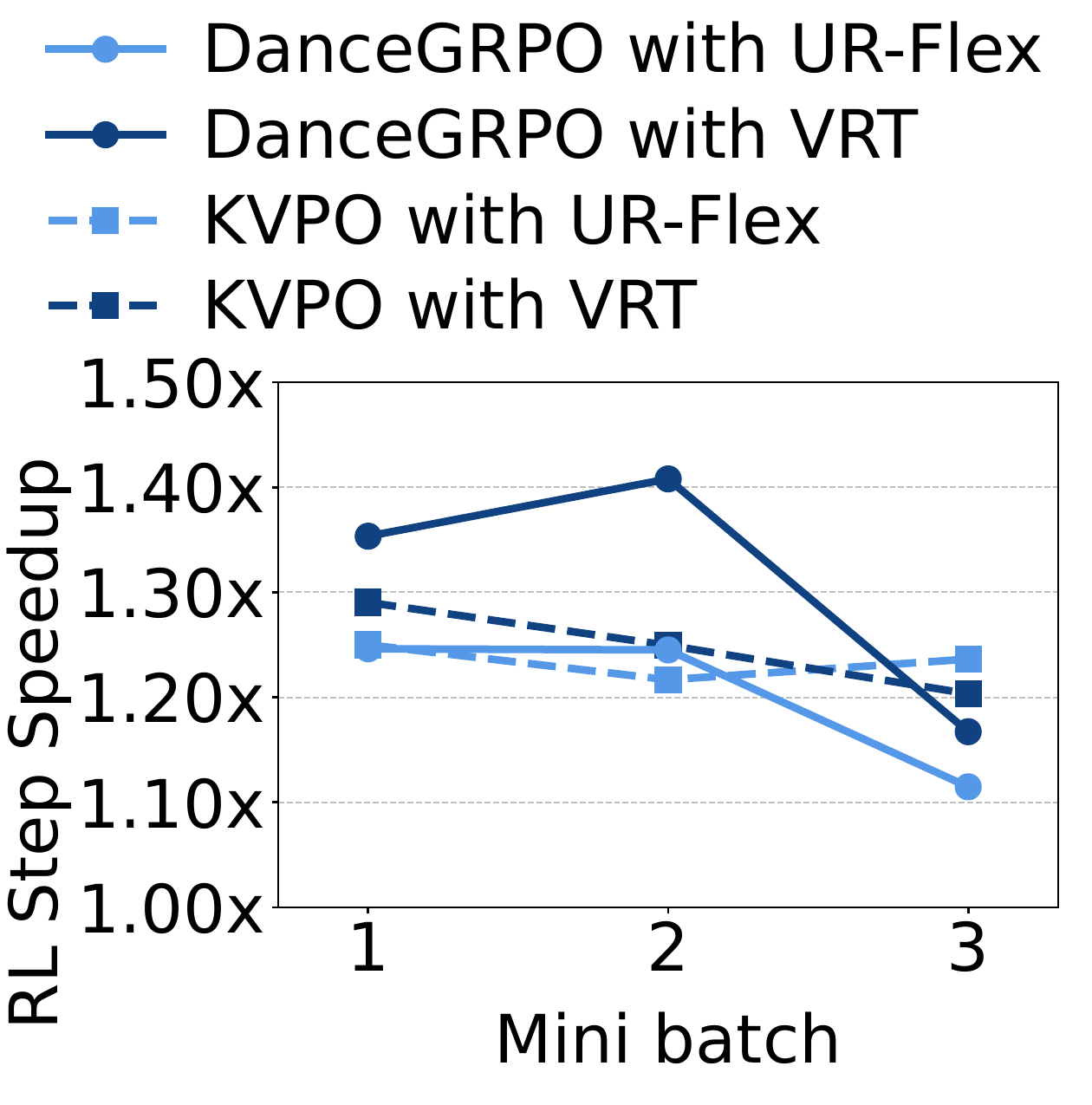}
    \vspace{-18pt}
    \caption{Number of mini-batches.}
    \vspace{-10pt}
    \label{fig:eval_minibatch}
  \end{minipage}\hfill%
  \begin{minipage}[t]{0.49\linewidth}
    \vspace{0pt}
    \centering
    \includegraphics[width=\linewidth]{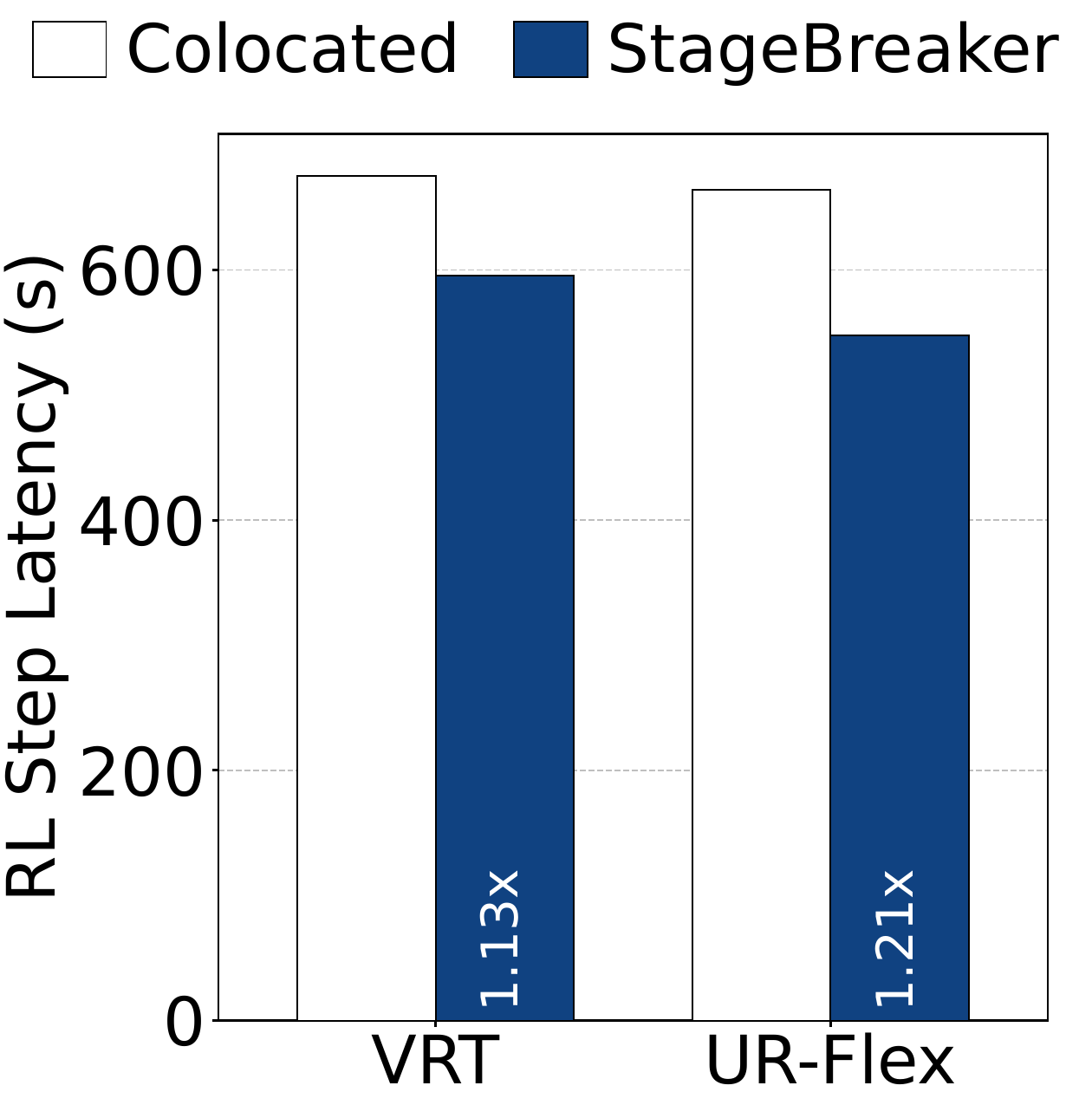}
    \vspace{-18pt}
    \caption{5B model testing.}
    \vspace{-10pt}
    \label{fig:eval_longlive5b}
  \end{minipage}
\end{figure}

\noindent \textbf{Impact of batch size.}
Figure~\ref{fig:eval_batch_size} varies the batch size from 6 to 24 candidates.
\texttt{StageBreaker} outperforms both baselines across all batch sizes, achieving a speedup of 1.12$\sim$1.41$\times$ over \texttt{Colocated} and 1.46$\sim$1.79$\times$ over \texttt{AsyncReward}.
Between the two RL algorithms, \texttt{StageBreaker} sustains a speedup of 1.19$\sim$1.41$\times$ with DanceGRPO and 1.12$\sim$1.31$\times$ with KVPO over the three batch sizes.
With DanceGRPO, the speedup further rises to 1.25$\sim$1.41$\times$ at 24 candidates, as a larger batch provides more computation tasks to relocate and thus balances the load between the actor and reward GPUs at a finer granularity.

\noindent \textbf{Impact of video resolution.}
Figure~\ref{fig:eval_resolution} compares \texttt{StageBreaker} with \texttt{Colocated} on DanceGRPO at 360p and 480p across the three video lengths. We omit KVPO here for brevity, as it exhibits a similar performance trend.
StageBreaker achieves a speedup of 1.13$\sim$1.30$\times$ at 360p and 1.05$\sim$1.32$\times$ at 480p.
The trend follows that of video length, as a higher resolution likewise raises the cost of rollout and policy training.
\texttt{StageBreaker} thus sustains a speedup of 1.17$\sim$1.32$\times$ at both resolutions with VR-Thinker, where the overlap of reward evaluation contributes most of the gain, and 1.13$\sim$1.24$\times$ at 360p and 1.05$\sim$1.19$\times$ at 480p with UnifiedReward-Flex.

\noindent \textbf{Impact of the number of mini-batches.}
Figure~\ref{fig:eval_minibatch} varies the number of mini-batches per RL step.
We use 24 candidates for DanceGRPO to ensure evenly sized mini-batches under its algorithmic constraints.
\texttt{StageBreaker} outperforms \texttt{Colocated} across all settings, achieving 1.11$\sim$1.41$\times$ speedup.
Between the two RL algorithms, its speedup stays stable with KVPO but varies with DanceGRPO.
Even as more mini-batches reduce the candidates in each mini-batch, KVPO still provides \texttt{StageBreaker} with enough separable computation tasks within policy training (e.g., the reference-policy scoring \cite{zhang2026kvpo}) to balance the GPU load.
DanceGRPO, in contrast, moves computation only as whole candidates and thus cannot partition each mini-batch finely enough to reduce the step latency under 3 mini-batches.

\noindent \textbf{Impact of model size.}
Figure~\ref{fig:eval_longlive5b} compares \texttt{StageBreaker} with \texttt{Colocated} on KVPO when scaling the video model from LongLive-1.3B to LongLive-5B \cite{yang2026longlive}.
Although larger models commonly adopt techniques such as FP4 acceleration in LongLive-5B to mitigate the video generation cost, these techniques speed up each denoising step alone and leave stage-centric execution intact.
\texttt{StageBreaker} instead targets the execution itself, and thus sustains a speedup of 1.13$\sim$1.21$\times$ on the larger model.

\subsection{Training Fidelity}
\begin{figure}[t]
  \centering
  \includegraphics[width=0.99\linewidth]{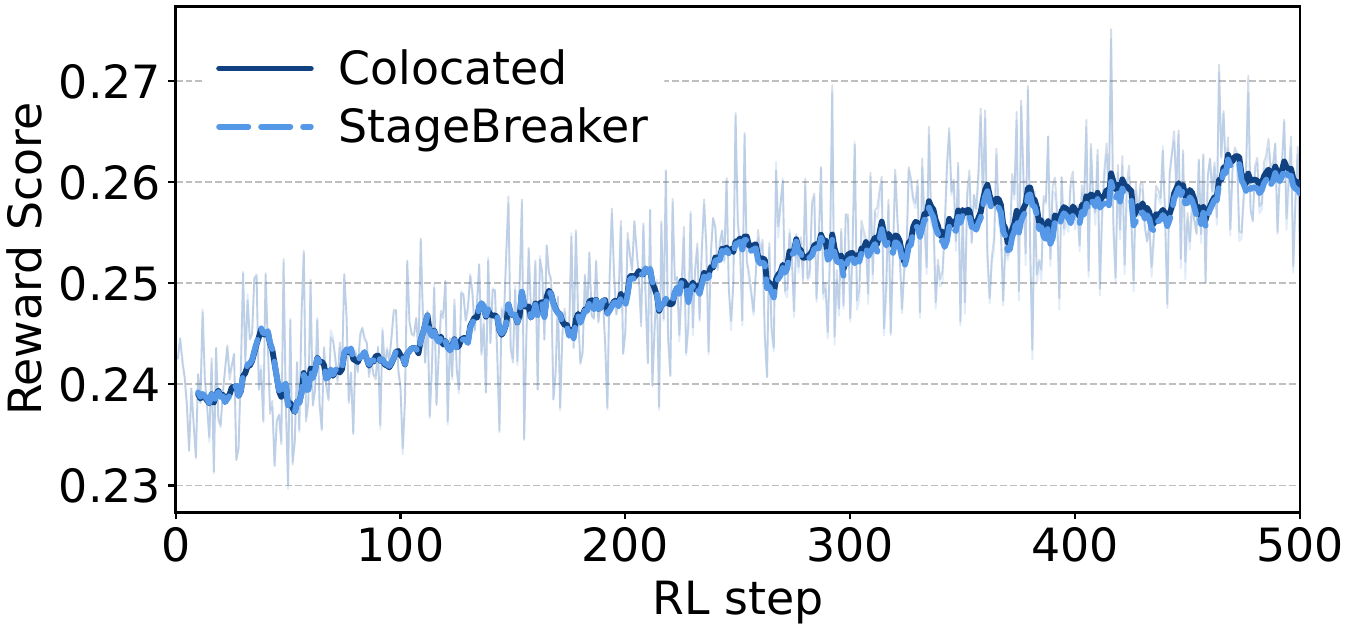}
  \vspace{-10pt}
  \caption{Reward curves.}
  \vspace{-10pt}
  \label{fig:fidelity}
\end{figure}

\texttt{StageBreaker} reschedules only the execution of an RL step and keeps the gradient of every weight update mathematically identical to that of stage-centric execution.
This identity holds in exact arithmetic, while floating-point accumulation on GPUs introduces small numerical differences \cite{zhuang2022randomness, summers2021nondeterminism}.
To verify that these differences do not alter training, we adopt the official recipe of DanceGRPO~\cite{xue2025dancegrpo,dancegrpo_github}, which trains Stable Diffusion 1.4~\cite{blattmann2023stable} with the HPS-v2.1~\cite{wu2023human}.
We run this RL post-training for 500 RL steps under \texttt{Colocated} and \texttt{StageBreaker}, and Figure~\ref{fig:fidelity} plots the training reward, the mean reward score of the candidates generated at each step (light lines), with its 10-step moving average (dark lines).
The two curves overlap throughout RL training without any systematic gap, both rising from about 0.24 to 0.26 over the 500 steps with only negligible differences at any step.
These differences stem from floating-point rounding on GPUs, which makes the same computation yield slightly different values when its operations run in a different order.
\texttt{StageBreaker} reorders these operations by applying the advantage and the importance ratio after back-propagation, but the resulting differences do not compromise the algorithmic fidelity of RL training.

\begin{figure}[t]
  \centering
  \includegraphics[width=0.99\linewidth]{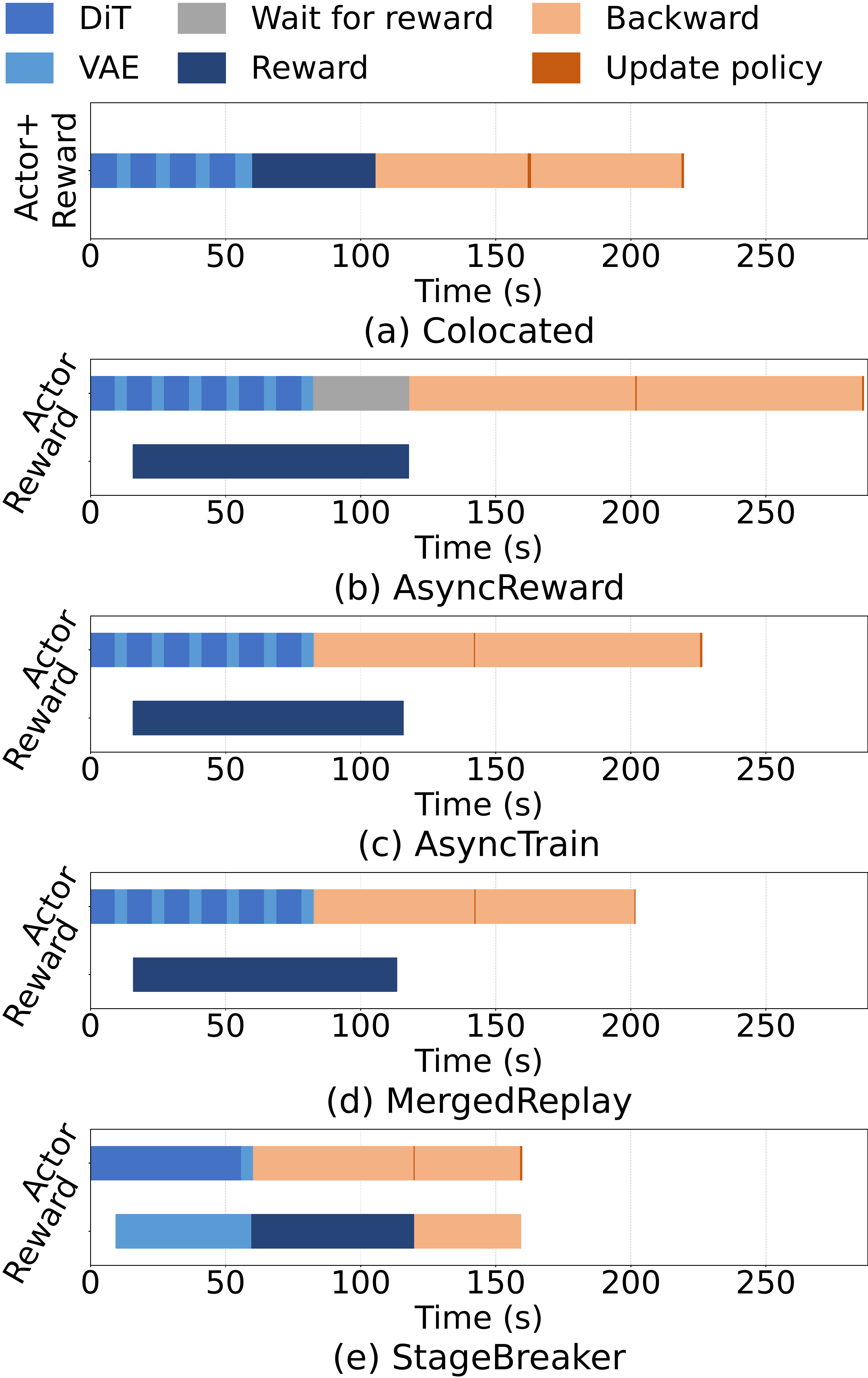}
  \vspace{-10pt}
  \caption{Execution timeline.}
  \vspace{-12pt}
  \label{fig:timeline_brk}
\end{figure}

\subsection{Performance Analysis}

\label{sec:eval-analysis}

Figure~\ref{fig:timeline_brk} traces the execution of an RL step on a representative actor GPU and reward GPU under the five compared systems, using DanceGRPO with VR-Thinker at a batch size of 24 candidates.

\noindent \textbf{Baseline.} \texttt{Colocated} runs the three stages in sequence on all six GPUs (cf. Figure~\ref{fig:timeline_brk}a), each of which generates four video candidates in rollout.
\texttt{AsyncReward} (cf. Figure~\ref{fig:timeline_brk}b) dedicates two of the six GPUs to the reward model, and the remaining four actor GPUs run rollout and policy training with six video candidates each.
The reward GPUs overlap 64\% of the scoring with rollout and hide this part of the reward cost from the critical path, while the remaining 36\% stays on it.
As the \texttt{Wait for reward} block in Figure~\ref{fig:timeline_brk}b shows, however, reward evaluation still blocks policy training for 13\% of the RL step after rollout, since policy training requires the complete reward scores of the entire batch.
Meanwhile, the four actor GPUs carry all rollout and policy training stages, whereas the two reward GPUs stay busy only for the 36\% of the RL step spent on scoring and sit idle for the remaining 64\%.
These challenges together waste 30\% of the GPU time of each RL step. \texttt{AsyncReward} thus reaches only 0.78$\times$ the performance of \texttt{Colocated}.

\noindent\textbf{StageBreaker.}
To address reward-blocked policy training, \texttt{AsyncTrain} runs back-propagation on the reward-independent part before the reward scores arrive.
As Figure~\ref{fig:timeline_brk}c shows, the actor GPUs start policy training immediately after rollout while the reward GPUs are still scoring, which removes the \texttt{Wait for reward} block of \texttt{AsyncReward} and reduces the RL step by 22\%.
\texttt{MergedReplay} then eliminates the redundant trajectory replay, running the DiT over each trajectory once with gradient tracking rather than twice (Figure~\ref{fig:timeline_brk}d), which cuts policy training on the actor GPUs by 17\% and achieves a speedup of 1.13$\times$ over \texttt{Colocated}.
The four actor GPUs, however, still carry every VAE decoding of rollout, which takes about 30\% of the rollout time, and every loss computation and back-propagation of policy training.
Meanwhile, the reward GPUs sit idle for 37\% of the RL step after the last score.
\texttt{StageBreaker} further addresses this stage-induced load imbalance with task-level load balancing.
As Figure~\ref{fig:timeline_brk}e shows, \texttt{StageBreaker} moves five VAE decodings to the reward GPUs under our placement plan, which shortens rollout on the actor GPUs by 27\%.
Scoring starts once these decodings finish, which gathers the scoring requests into a larger batch and thus accelerates the memory-bound CoT decoding of the VLM.
After scoring, the reward GPUs further take over a portion of the loss computation and back-propagation, finishing policy training 15\% earlier.
The idle bubbles on the reward GPUs thus mostly disappear, raising their utilization from 36\% under \texttt{AsyncReward} to 84\%.
\texttt{StageBreaker} thereby achieves a speedup of 1.41$\times$ over \texttt{Colocated} and 1.81$\times$ over \texttt{AsyncReward}.

\section{Related Works and Discussions}

\label{sec:related}

\noindent\textbf{Video generation.}
Numerous efforts have built DiT-based video generation models~\cite{wu2025hunyuanvideo,wan2025wan,teng2025magi,chen2025skyreels,yin2025slow,huang2025selfforcing,yang2026longlive,liu2026rolling,cui2026self,chen2026longlive}.
Bidirectional models such as HunyuanVideo~\cite{wu2025hunyuanvideo}, Wan~\cite{wan2025wan}, CogVideoX~\cite{yang2025cogvideox}, and Open-Sora~\cite{zheng2024open} attend over all video frames at once, which achieves high visual quality but confines generation to short clips of fixed length and precludes streaming.
To overcome these limitations, recent works turn to AR models~\cite{teng2025magi,chen2025skyreels,agarwal2026cosmos}, which instead generate videos chunk by chunk under causal attention, supporting long-form and streaming generation.
To further reduce the generation cost, CausVid~\cite{yin2025slow}, Self-Forcing~\cite{huang2025selfforcing}, and their successors~\cite{yang2026longlive,liu2026rolling,cui2026self,chen2026longlive} distill few-step AR generators from bidirectional models, cutting the denoising steps of each chunk from tens to a few (e.g., from 50 to 4) and thus achieving real-time long video generation.
StageBreaker readily accommodates all these AR models, as it reschedules only the execution of the RL step and modifies neither their architectures nor generation procedures.

\noindent\textbf{RL for video generation.}
Numerous works~\cite{xue2025dancegrpo,liu2025flow,zhang2026kvpo,lu2026raven,zhang2026astrolabe,he2026ar,wang2026tempact} post-train video models with RL to improve human preference alignment, motion quality, and temporal coherence.
DanceGRPO~\cite{xue2025dancegrpo} and Flow-GRPO~\cite{liu2025flow} extend GRPO to video diffusion and flow-matching models through SDE sampling.
Building on this general formulation, KVPO~\cite{zhang2026kvpo}, CM-GRPO~\cite{lu2026raven}, and other recent works~\cite{he2026ar, zhang2026astrolabe, wang2026tempact} further customize exploration and the surrogate policy for the few-step AR models.
To provide these algorithms with more accurate rewards, numerous works propose VLM-based reward models \cite{liu2025improving,wang2026unified,wang2025vr,wang2026unifiedThink,he2025videoscore2,wang2024lift,wang2026cac}, such as VideoAlign~\cite{liu2025improving}, UnifiedReward-Flex~\cite{wang2026unified}, and VR-Thinker~\cite{wang2025vr}, which replace lightweight metrics with multi-dimensional judgments through CoT reasoning.
These works improve RL post-training and are orthogonal to StageBreaker, which accelerates its execution without modifying the algorithms or reward models.

\noindent\textbf{RL post-training systems.}
A growing number of systems accelerate RL post-training of LLMs by orchestrating its stages across GPUs~\cite{sheng2025hybridflow,hu2025openrlhf,zhong2025optimizing,zhu2026distrs,fu2026areal,zhong2025streamrl,gao2026rollpacker}.
Frameworks such as HybridFlow~\cite{sheng2025hybridflow} and OpenRLHF~\cite{hu2025openrlhf} execute each RL step stage by stage over the entire batch, while RLHFuse~\cite{zhong2025optimizing} and RLinf~\cite{yu2026rlinf} decompose the stages into sample-level subtasks and micro-batches and recompose them across GPUs to fill the bubbles between stages.
To further shorten the critical path, StreamRL~\cite{zhong2025streamrl} and AReaL~\cite{fu2026areal} turn to off-policy training, which overlaps rollout with policy training on stale rollouts at the cost of the on-policy semantics~\cite{gao2026rollpacker,qin2026seer}.
RollPacker~\cite{gao2026rollpacker} and DistRS~\cite{zhu2026distrs} keep these semantics and overlap only reward evaluation with rollout.
All of them, however, back-propagate a sample only after its advantage arrives.
StageBreaker instead back-propagates ahead of the reward scores and applies the advantage afterwards, which overlaps policy training with reward evaluation while preserving on-policy semantics exactly.
Moreover, these systems optimize LLM rollout, whose memory-bound decoding and long-tail output lengths motivate batching and tail-aware scheduling~\cite{gao2026rollpacker,zhong2025streamrl,fu2026areal,qin2026seer}.
Video rollout, in contrast, gains little from batching and runs a fixed number of denoising steps for every video, which raises neither issue.

\noindent\textbf{RL systems for diffusion and video.}
Systems for diffusion RL mostly accelerate rollout, for example by offloading seed exploration to spot GPUs~\cite{lai2026spotlight} or sharing generation prefixes across candidates~\cite{li2026branchgrpo}.
DigenRL~\cite{wang2026accelerating} further overlaps rollout with training and lets the idle trainer assist generation, but still back-propagates after the scores arrive.
LeanGRPO~\cite{wang2026leangrpo}, a concurrent work, removes the redundant trajectory replay by back-propagating during rollout itself.
It applies, however, only when rollout and training run on the same engine under the same policy, a narrow setting for current RL systems.
In contrast, StageBreaker merges the two replays into one only on the training engine under the current policy, which enables StageBreaker to accommodate separate rollout and training engines as well as every policy update.
More importantly, StageBreaker observes the rising cost of VLM-based reward models, which brings new challenges to stage-centric execution overlooked by these systems built around lightweight rewards.
StageBreaker breaks stage-centric execution into a dataflow of computation tasks and schedules them by their true data dependencies to address these challenges.

\noindent\textbf{Dependency-driven scheduling.}
A wide range of ML systems schedule computation by its data dependencies rather than by program boundaries \cite{moritz2018ray,qi2024zerobubble,huang2019gpipe,narayanan2019pipedream,abadi2016tensorflow,barham2022pathways}.
Ray~\cite{moritz2018ray} runs a dynamic task graph in which each task starts once its inputs arrive, and zero-bubble pipeline parallelism~\cite{qi2024zerobubble} further splits the backward pass into input-gradient and weight-gradient computations to fill the remaining bubbles.
StageBreaker applies the same principle to video RL post-training, scheduling the true dataflow of an RL step, in which the loss and its gradient separate into independent factors and each stage decomposes into individually placeable tasks.

\section{Conclusion}

This paper presents StageBreaker, a data-dependency-driven scheduling framework for AR video RL post-training in resource-constrained settings, where a small and fixed GPU budget hosts rollout, VLM-based reward evaluation, and policy training.
StageBreaker overlaps policy training with reward evaluation, merges the two redundant trajectory replays, and balances the load across GPUs through task-level placement.
Evaluation results show that it achieves consistent speedups over stage-centric execution and asynchronous reward evaluation.
In the future, we plan to explore further RL optimizations in resource-constrained settings, including placement across heterogeneous hardware devices and scheduling based on their performance characteristics.

\appendix

\section*{Appendix}

\section{Back-Propagation through a DiT Block}
\label{app:backward}

We verify that back-propagation through a typical DiT block~\cite{wan2025wan,wu2025hunyuanvideo} preserves $A_i$ as a scalar factor, as established in \S\ref{sec:async-train}.
To this end, we show that each operation's Jacobians are independent of $A_i$ and its backward computation is linear in the upstream gradient.

The DiT block applies self-attention over the current chunk and the KV cache, cross-attention to the prompt embedding $E_i$, and an FFN.
Each sublayer operates on normalized hidden states, and its output is added to the residual stream.
\begin{align}
  X_1 &= X + \mathrm{Attn}\big(\hat{X}W_Q, [K^{<c};\hat{X}W_K], [V^{<c};\hat{X}W_V]\big)\,W_O, \label{eq:app-self}\\
  X_2 &= X_1 + \mathrm{Attn}\big(\hat{X}_1W_{Q'}, E_iW_{K'}, E_iW_{V'}\big)\,W_{O'}, \label{eq:app-cross}\\
  X_3 &= X_2 + \phi\big(\hat{X}_2W_1\big)\,W_2, \label{eq:app-ffn}
\end{align}
where $X\in\mathbb{R}^{L\times d}$ holds the $L$ tokens of the current chunk, $\hat{X} = \mathrm{LN}(X)$, $K^{<c}$ and $V^{<c}$ are the cached keys and values of the preceding chunks, and $\phi$ is GELU.
We omit the explicit formulas for timestep modulation and RoPE as they are simple affine and linear transformations, respectively.
Their backward computations multiply the upstream gradient by matrices independent of $A_i$, preserving $A_i$ as a scalar factor.

We examine the attention sublayers and the FFN, then the normalization and residual operations shared by all three sublayers.
For an unclipped loss term $\psi = A_i\rho_{i,k}$, let $\bar{z} = \partial\psi/\partial z$ denote the gradient with respect to $z$.

The attention sublayers in Equations~\ref{eq:app-self} and~\ref{eq:app-cross} combine linear projections with scaled dot-product attention.
For a linear projection $Y = XW$, the gradients are $\bar{X} = \bar{Y}W^{\top}$ and $\bar{W} = X^{\top}\bar{Y}$.
Under LoRA, $W = W_0 + DB$, the trainable parameter gradients are $\bar{D} = X^{\top}\bar{Y}B^{\top}$ and $\bar{B} = D^{\top}X^{\top}\bar{Y}$.
These computations are linear in $\bar{Y}$, with coefficients independent of $A_i$. For scaled dot-product attention, each head computes $S = QK^{\top}/\sqrt{d_h}$, $P = \mathrm{softmax}(S)$, and $O = PV$.
Here, $d_h$ is the head dimension, and softmax is applied row-wise.
The backward equations are
\begin{align}
  \bar{V} &= P^{\top}\bar{O},\qquad \bar{P} = \bar{O}V^{\top}, \nonumber\\
  \bar{S} &= P\odot\bar{P} - \mathrm{diag}\big((P\odot\bar{P})\mathbf{1}\big)\,P, \label{eq:app-attn}\\
  \bar{Q} &= \bar{S}K/\sqrt{d_h},\qquad \bar{K} = \bar{S}^{\top}Q/\sqrt{d_h}. \nonumber
\end{align}
Each equation is linear in its upstream gradient, while its coefficients are determined by the forward-pass quantities $P$, $Q$, $K$, and $V$, all independent of $A_i$.
These equations apply to both self-attention and cross-attention and remain valid under attention masking.
Together with the linear projections, they preserve $A_i$ as a scalar factor throughout each attention sublayer.

The FFN in Equation~\ref{eq:app-ffn} consists of two linear projections and a GELU activation.
The projections follow the formulas above.
For the activation $Y = \phi(X)$, the backward computation is $\bar{X} = \phi'(X)\odot\bar{Y}$.
Since $\phi'(X)$ depends only on the forward input, this computation also preserves $A_i$ as a scalar factor.

All three sublayers share normalization and residual operations.
For the normalization step of LayerNorm, let $n$ be the normalized vector for token $x$ and $\sigma$ the normalization scale.
Its backward computation is
\[
  \bar{x} = \frac{1}{\sigma}
  \left(I - \frac{1}{d}\mathbf{1}\mathbf{1}^{\top}
  - \frac{1}{d}nn^{\top}\right)\bar{n}.
\]
This computation is linear in $\bar{n}$, while its coefficients depend only on the forward-pass values $n$ and $\sigma$, both independent of $A_i$.
A residual addition passes the upstream gradient to both inputs.
When a hidden state feeds multiple branches, Equation~\ref{eq:vjp} sums their gradients, also preserving the common factor $A_i$.

For an upstream gradient $\bar{y} = A_i h$ with $h$ independent of $A_i$, each operation therefore preserves $A_i$ as a scalar factor in both $\bar{x}$ and $\bar{W}$.
This property holds across all three sublayers and across stacked blocks, so $A_i$ remains a scalar factor throughout back-propagation through the entire DiT.

\bibliographystyle{ACM-Reference-Format}
\bibliography{references}

\end{document}